\documentclass[
prd,
11pt,
onecolumn,
preprintnumbers,
superscriptaddress,
tightenlines,
nofootinbib,
letterpaper,
longbibliography,
twoside
]{revtex4-2}

\usepackage{multirow}
\usepackage{epsfig}
\usepackage{graphicx}
\usepackage{subfigure}
\usepackage{placeins}
\usepackage{mathrsfs}
\usepackage{amssymb}
\usepackage{amsmath}
\usepackage{float}
\usepackage{verbatim}
\usepackage{enumitem}
\usepackage[toc,page]{appendix}
\usepackage{lmodern}
\usepackage{url}
\usepackage[percent]{overpic}
\usepackage{orcidlink}
\usepackage[normalem]{ulem}
\usepackage[table,xcdraw,dvipsnames]{xcolor}
\usepackage{colortbl}

\definecolor{MyDarkBlue}{rgb}{0.1,0.1,0.8}
\definecolor{SBlue}{rgb}{0.2,0.4,0.7}
\definecolor{MyLightBlue}{rgb}{0.22,0.51,0.9}
\definecolor{MyGreen}{rgb}{0.0,0.5,0.0}
\definecolor{BrickRed}{rgb}{0.8,0.25,0.33}

\usepackage{hyperref}

\hypersetup{
  colorlinks=true,
  citecolor=SBlue,
  linkcolor=MyDarkBlue,
  urlcolor=PineGreen
}

\begin{document}

\title{Impact of Bubble Nucleation History and Friction on Primordial Black Hole Formation}

\author{Tathagata Ghosh\,\orcidlink{0000-0002-8259-0328}}
\email{tathagataghosh@hri.res.in}
\affiliation{Regional Centre for Accelerator based Particle Physics, Harish-Chandra Research Institute, Chhatnag Road, Jhunsi, Prayagraj 211019, India}
\affiliation{Homi Bhabha National Institute, Training School Complex, Anushakti Nagar, Mumbai
400094, India}
\author{Kousik Loho\,\orcidlink{0000-0001-6330-9505}}
\email{kousikloho@hri.res.in}
\affiliation{Regional Centre for Accelerator based Particle Physics, Harish-Chandra Research Institute, Chhatnag Road, Jhunsi, Prayagraj 211019, India}
\author{Sudip Manna\,\orcidlink{0009-0000-2162-0506}}
\email{sudipmanna@hri.res.in}
\affiliation{Regional Centre for Accelerator based Particle Physics, Harish-Chandra Research Institute, Chhatnag Road, Jhunsi, Prayagraj 211019, India}
\affiliation{Homi Bhabha National Institute, Training School Complex, Anushakti Nagar, Mumbai
400094, India}

\begin{abstract}
Cosmological first-order phase transitions (FOPTs) in the early Universe are an exciting prediction of many beyond the Standard Model scenarios. Owing to the stochastic nature of bubble nucleation, some causal patches may remain trapped in the false vacuum long after the surrounding regions have completed the transition and started to redshift. Under suitable conditions, these delayed patches can become sufficiently overdense to collapse into primordial black holes (PBHs). One therefore expects characteristic parameters, such as the PBH mass and PBH formation time to be sensitive to the underlying phase-transition dynamics. Here, we adopt a model-independent framework with a fixed mean bubble separation scale ($R_*$) to directly compare the impact of exponential and Gaussian nucleation profiles on PBH formation. We further incorporate the effects of friction on the bubble walls arising from interactions with the surrounding plasma. We find that friction significantly modifies the bubble dynamics, leading to significant changes in the PBH mass and formation time for both nucleation histories. Moreover, the stochastic gravitational-wave spectrum generated by the FOPT can provide a complementary probe of the underlying dynamics, allowing the effects of friction on the bubble wall evolution to be distinguished from the frictionless case.

\end{abstract}
\begin{flushright}
\small{HRI-RECAPP-2026-08}
\end{flushright}
\maketitle

\tableofcontents
\newpage
\section{Introduction}
\label{sec:intro}

Measurements of the cosmic microwave background (CMB) anisotropies have provided a unique observational window onto the early Universe, particularly the epoch around recombination.~\cite{WMAP:2012nax,Planck:2018vyg}. However, our understanding of the thermal history of the Universe prior to recombination remains limited, as direct electromagnetic observations cannot probe epochs earlier than the surface of last scattering. The primordial abundances of the light elements inferred from astrophysical observations in conjunction with the CMB data can provide an important probe of an epoch only as far back as Big Bang Nucleosynthesis (BBN) \cite{Cyburt:2015mya,Pitrou:2018cgg}. Consequently, obtaining observational evidence of the pre-BBN Universe has remained one of the major challenges of modern cosmology.

Gravitational waves (GWs) have emerged as a very important new observational handle on the early Universe, offering a unique opportunity to probe otherwise inaccessible pre-BBN epochs. A major breakthrough has come in GW astronomy with the detection of binary black hole mergers~\cite{LIGOScientific:2016aoc} as well as binary neutron star inspirals~\cite{LIGOScientific:2017vwq}. 
These detections established GW astronomy as a new observational field and have motivated extensive efforts to detect a stochastic GW background, including a possible cosmological component that could probe the pre-BBN Universe.
In particular, the recent evidence of a stochastic GW background by pulsar timing array (PTA) collaborations has attracted significant attention~\cite{NANOGrav:2023gor,NANOGrav:2023hde,EPTA:2023fyk,Reardon:2023gzh,Xu:2023wog}. 
Possible cosmological interpretations of the NANOGrav 15-year (NG15) data have subsequently been extensively discussed, including first-order phase transitions in the early Universe, cosmic strings, domain walls and other sources~\cite{NANOGrav:2023hvm}.
Although none of these cosmological interpretations is firmly established, the future identification of a stochastic GW background of cosmological origin would provide a promising avenue for probing the physics of the pre-BBN Universe.

Another cosmological relic that has attracted significant attention in recent years is the primordial black hole (PBH). Besides being a viable dark matter candidate over certain mass ranges, PBHs are associated with a variety of observable signatures (including gravitational waves \cite{Saito:2008jc,Inomata:2020lmk}) making them an important probe of the early Universe. Since the inception of PBH ~\cite{Zeldovich:1967lct,Hawking:1971ei} various interesting mechanisms have been suggested for their formation. These include the collapse of large primordial density perturbations~\cite{Zeldovich:1967lct,Hawking:1971ei,Carr:1974nx,Carr:1975qj}, collapse of cosmic strings~\cite{Hawking:1987bn,Polnarev:1988dh,Jenkins:2020ctp}, collapse of domain walls~\cite{Deng:2016vzb,Garriga:2015fdk,Liu:2019lul,Zeldovich:1974uw,Ge:2019ihf}, collision of bubble walls during FOPTs~\cite{Crawford:1982yz,Kodama:1982sf,Hawking:1982ga,Moss:1994pi,Freivogel:2007fx}, delayed FOPTs~\cite{Jedamzik:1999am,Liu:2021svg,Gouttenoire:2023naa,Lewicki:2023ioy,Kanemura:2024pae}, Q-ball-induced fluctuations~\cite{Cotner:2016cvr}, collapse of Fermi-balls~\cite{Kawana:2021tde}, and several other phenomena associated with FOPTs~\cite{Baker:2021nyl,Baker:2021sno}. In this work, we focus on PBH formation through the collapse of overdensities generated during delayed FOPTs.

Among the cosmological sources discussed above in the context of PTA data, cosmological FOPTs are a tantalizing prediction of many beyond the Standard Model (BSM) scenarios. Such a transition proceeds from a metastable false vacuum to the true vacuum through bubble nucleation, where the two vacua are separated by a potential barrier. Bubble nucleation is an inherently stochastic process, implying that different regions of the Universe undergo the phase transition at different times. Regions in space that have already transitioned to the true vacuum start nucleating expanding bubbles of true vacuum. As these bubbles grow, collide and eventually percolate, the phase transition undergoes completion throughout the Universe.

Owing to the stochastic nature of bubble nucleation, certain causally connected regions may remain trapped in the false vacuum significantly longer than the surrounding regions. Such delayed bubble nucleation leads to delayed percolation, where bubbles within the delayed patch nucleate much later than those in the surrounding Universe. Consequently, while the surrounding regions have already completed the phase transition and their energy density has begun to redshift as radiation, the delayed patch continues to retain the false vacuum energy. When the phase transition eventually completes within the delayed patch, a local overdensity (in energy) is generated and under suitable conditions this overdensity can gravitationally collapse to form a PBH. The resulting PBH mass and formation time are therefore highly sensitive to the underlying dynamics of the FOPT.

The dynamics of cosmological FOPTs are influenced by several physical effects. One important ingredient is the friction experienced by the expanding bubble walls due to their interactions with the surrounding thermal plasma. Bubble-wall friction affects the wall velocity, the completion of the phase transition, and consequently the evolution of the delayed patches responsible for PBH formation. Another important aspect of the delayed FOPT and PBH formation is the bubble nucleation history. While different particle-physics models can exhibit either exponentially growing or approximately Gaussian bubble nucleation rates, the impact of these distinct nucleation histories on PBH formation has not been been systematically compared within a common and model-independent framework. Since both bubble-wall friction and the nucleation history directly influence the phase-transition dynamics, they are expected to have a significant impact on PBH formation.

In this work, we present a model-independent analysis of PBH formation from delayed FOPTs by comparing exponential and Gaussian nucleation histories while fixing the mean bubble separation scale, $R_*$. We consistently incorporate bubble-wall friction into the dynamics of the expanding bubbles and investigate its impact on bubble expansion, the characteristic PBH mass, and the PBH formation time. Our results demonstrate that the combined observations of PBHs and stochastic GWs provide complementary probes of the underlying phase-transition dynamics and the mechanism responsible for PBH formation.

The paper is organized as follows. In Sec.~\ref{sec:DPT}, we briefly introduce the mechanism of delayed FOPTs and discuss the effects of bubble-wall friction and the bubble nucleation history on PBH formation. The stochastic GW signal from such phase transitions is presented in Sec.~\ref{sec:GW}. Finally, we conclude in Sec.~\ref{sec:conc}.


\section{Primordial Black Holes from Delayed First-Order Phase Transition}
\label{sec:DPT}

Cosmological FOPT is a process that describes the drastic transition of the equation of state parameter ($w$) of the Universe from the vacuum dominated Universe ($w=-1$) to the radiation domination ($w=1/3$). The latent heat required for the phase transition is sourced from energy density associated with the vacuum. Due to the statistical nature of this process the bubble nucleation time may differ leading to distinct percolation times and the formation of delayed causal patches. It has to be noted that while the energy density dominated by the vacuum remains constant in those causal patches, the surrounding region has already transitioned to the radiation dominated region and thus the energy density dilutes with the scale factor as $\rho_{R}\sim a^{-4}$. Finally, when the percolation takes place in the delayed patch and the energy density contained in the vacuum is converted to radiation, an overdensity is created. The measure of the overdensity is parametrized as $\delta$, given by
\begin{equation}
\delta=\frac{\rho_R^{dp}-\rho_R^{bkg}}{\rho_R^{bkg}},
\label{eq:delta}
\end{equation}
where $\rho_R^i$ signifies the radiation energy densities of the delayed patch (dp) and the surrounding background (bkg). If this overdensity reaches a critical value $\delta_c$, it can trigger a gravitational collapse in the causally connected delayed patch ultimately leading to the formation of a PBH. 

In the post-inflationary and pre-BBN era, the Universe might reside in a false vacuum and thus the energy density of the vacuum might start to dominate over the radiation energy density. In order to describe the framework, we define a starting point that has a temperature $T_{eq}$ at time $t_{eq}$ when the energy densities of radiation and vacuum becomes equal followed by a vacuum dominated epoch. The vacuum energy density at this temperature ($\rho_V(T_{eq})$) is given by
\begin{equation}
\rho_V(T_{eq})=\Delta V=\rho_{R}(T_{eq})=\frac{\pi^2}{30}g_\star(T_{eq})T_{eq}^4,
\label{eq:delv}
\end{equation}
where $\Delta V$ is difference between the false and true vacuum energy densities. The $T_{eq}$ is related to the nucleation temperature $T_n$ through the latent heat parameter
\begin{equation}
\alpha\sim\frac{\rho_V}{\rho_R}\bigg|_{T=T_n}\approx\bigg(\frac{T_{eq}}{T_n}\bigg)^4,
\label{eq:alpha}
\end{equation}
which is a measure of the strength of the FOPT. The duration of the FOPT is measured in terms of the $\beta^{-1}$ parameter in the case of an exponential bubble nucleation rate.

The exponential (i.e., with an exponent linear in time) bubble nucleation rate per unit volume $\Gamma_V$ is given by \cite{Gouttenoire:2023naa}
\begin{equation}
\Gamma_V (t)=H(T_n)^4e^{\beta(t-t_n)},
\label{eq:gamma}
\end{equation}
where $t_n$ is the time instance of nucleation and $H(T_n)$ is the hubble parameter at temperature $T_n$. For the rest of the text we refer to this exponent linear in time as the ``exponential" bubble nucleation rate. The evolution of the radiation energy density follows the equation
\begin{equation}
\frac{d\rho_R^i}{dt}+4H\rho_R^i=-\frac{d\rho_V^i}{dt},
\label{eq:evol}
\end{equation}
where the hubble parameter is given by $H^i=\sqrt{\frac{\rho_R^i+\rho_V^i}{3M_{pl}^2}}$ and the vacuum energy density $\rho_V^i$ by
\begin{equation}
\rho_V^i(t)=\Delta V\exp\bigg[-\int_{t_n}^td\tilde t \Gamma_V^i(\tilde t)a^i(\tilde t)^3\frac{4}{3}\pi R^i(\tilde t)^3\bigg],
\label{eq:rhoV}
\end{equation}
where $a^i(t)$ and $R^i(t)$ are the scale factor and the comoving bubble radius at time $t$ respectively. The scale factor and the comoving radius evolve according to the respective equations given by
\begin{equation}
\frac{da^i}{dt}=a^iH^i,
\label{eq:sf}
\end{equation}
and
\begin{equation}
R^i(t)=\int_{t_n}^t\frac{d\tilde t}{a^i(\tilde t)}.
\label{eq:radius}
\end{equation}
The index ``$i$" is indicative of the fact that all of the indexed quantities exist in pairs - one for the delayed patch (dp) and the other for the background (bkg). From the above equations, it is evident that the quantities such as the energy densities, scale factor, comoving bubble radius, nucleation rate etc. depend on the nucleation time $t_n$.

Due to the inherent probabilistic nature of FOPTs, the bubble nucleation may not take place simultaneously in spatially separated regions leading to the formation of delayed patches \cite{Kodama:1982sf,Hashino:2021qoq,Gouttenoire:2023naa}. In such patches the bubble nucleation is delayed and hence, the bubble nucleation time is greater than the surrounding region which is referred as the background and the corresponding nucleation time of the background is denoted by $t_n^{bkg}=t_n$. This delay in bubble nucleation ultimately results in delayed transition in that patch and an over dense region is  created. We define a nucleation time $t_n^{dp}=t_{PBH}$ such that the patch is delayed enough to generate the required overdensity in order for it to gravitationally collapse and form a black hole. It goes without saying that $t_{PBH}>t_n$.

\begin{figure}[htb!]
\centering
\includegraphics[scale=0.5]{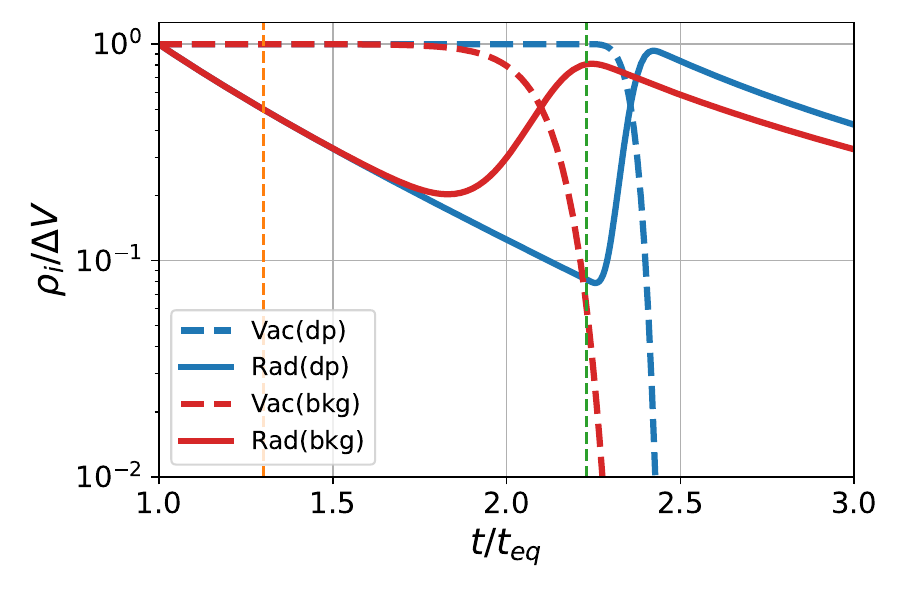}
\includegraphics[scale=0.5]{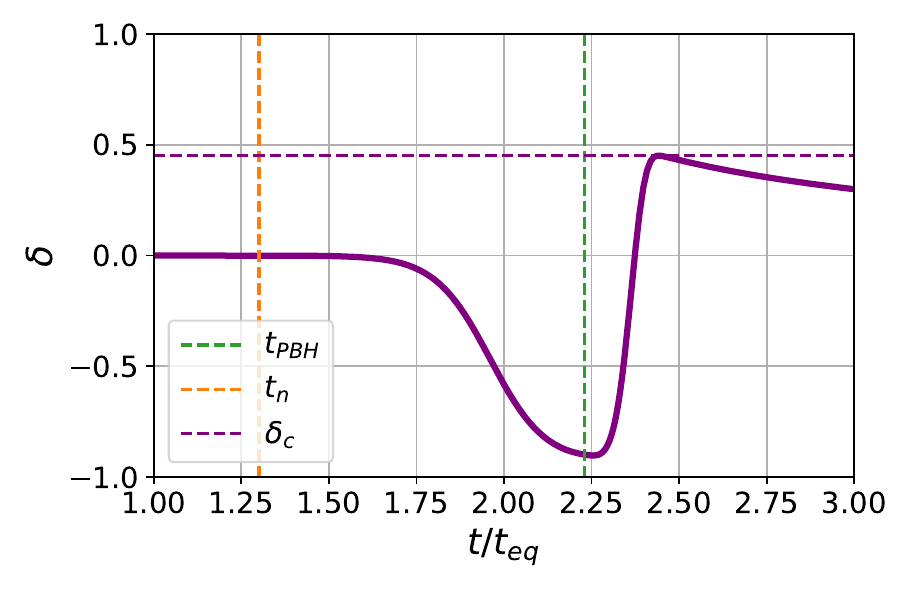}
\caption{\textit{Left:} The evolution of energy densities of vacuum and radiation for background as well as the delayed phase with time for an exponential nucleation rate (Eq.~\eqref{eq:gamma}) without taking into account the effect of friction. The benchmark values used are $\alpha=2$ and $\beta/H_n=20$ which provides $t_n=1.300\times t_{eq}$. The minimum delay required to produce PBH is given by $t_{PBH}=2.230\times t_{eq}$. \textit{Right:} Evolution of the $\delta$ parameter that determines PBH formation. This corresponds to BP 1 on Table~\ref{tab:BP}.}
\label{fig:evol_exp_gamma_0}
\end{figure}

We solve the aforementioned coupled differential equations numerically\footnote{Some nuances of numerically solving the equations are mentioned in Ref.~\cite{Banerjee:2024cwv}.} to arrive at Fig.~\ref{fig:evol_exp_gamma_0}. The value of $\delta_c$ can span the range $0.40\lesssim\delta_c\lesssim0.67$ depending upon the shape of the density profile \cite{Musco:2018rwt,Escriva:2019phb,Ning:2026nfs} (for a discussion on the gauge dependencies see Refs.~\cite{Franciolini:2025ztf,Harada:2015yda,Ai:2026zrs}). However, we have chosen the value of $\delta_c$ to be equal to 0.45, as has been widely used in the literature \cite{Jedamzik:1999am,Green:2004wb,Musco:2004ak,Harada:2015yda}. We do not expect any qualitative change in the central theme of this work with a small shift in the $\delta_c$ value. The gravitational collapse of such an overdensity can lead to the formation of a PBH with its initial mass given by

\begin{equation}
M_{BH}^{in}=\frac{4}{3}\pi\gamma\frac{\rho_R(T_{in})}{H({T_{in}})^3}\sim\bigg(\rho_R(T_{in})\bigg)^{-1/2},
\label{eq:BHmass}
\end{equation}

where the $\gamma$ is an efficiency parameter of collapse which takes a value 0.2 approximately if the collapse happens during radiation domination and we define the temperature of the collapse as $T_{in}$. While we assume for simplicity that the PBHs follow a monochromatic mass profile, in a broader mass distribution Eq.~\eqref{eq:BHmass} represents the peak mass\footnote{A study with PBH mass distributions can be found in Refs.~\cite{Lewicki:2024ghw,Lewicki:2024sfw}.}. In Ref.~\cite{Dutta:2026pbm}, the Schwarzschild collapse criterion for PBH formation is generated dynamically without using an overdensity threshold and the corresponding GW signal is discussed.

The evolution of the vacuum and radiation energy densities along with the corresponding overdensity parameter $\delta$ relevant for PBH formation, is shown in Fig.~\ref{fig:evol_exp_gamma_0}. On the left panel, the energy densities corresponding to the delayed patches are drawn in blue colour while the ones corresponding to the background are in red. The vacuum energy densities are shown using dashed lines and the radiation energy densities in solid lines. The overdensity parameter is shown using a solid purple curve on the right panel and the critical value is denoted using a dashed purple line. The results correspond to an exponential nucleation rate without including the contribution from the friction exerted on the bubble walls due to their interactions with the surrounding plasma. At early times, around $t_{eq}$ (before $t_n$), both the background and the delayed patches reside in the false vacuum. As a consequence, the vacuum energy densities in both regions remain nearly constant, while the radiation energy densities gradually start diluting as $a^{-4}$ due to the expansion of the Universe. Since both regions evolve almost identically during this stage, no significant density contrast is generated and the overdensity parameter $\delta$ remains negligible ($\delta \approx 0$).

As the Universe evolves, bubble nucleation first becomes efficient in the background around $t_n$, initiating the conversion of false vacuum energy into radiation in the background region. Therefore, the background vacuum energy density gradually decreases, while the corresponding background radiation energy density starts to increase. In contrast, the delayed patch still remains in the false vacuum, so its vacuum energy density is almost unchanged and its radiation component continues to redshift. As a result, the background becomes temporarily denser in radiation than the delayed patch, causing the overdensity parameter to decrease and even become negative.

At later times, around $t_{\rm PBH}$, bubble nucleation also becomes efficient inside the delayed patch. The delayed vacuum energy is then rapidly converted into radiation, producing a much larger radiation energy density than in the surrounding background, since the phase transition has already completed there and the released radiation has already started redshifting. This delayed release of latent heat in the delayed patches generates a significant positive density contrast between the delayed patch and the background. Consequently, the overdensity grows rapidly and can exceed the threshold required for gravitational collapse, leading to the formation of PBHs.

The quantities shown in Figs.~\ref{fig:evol_exp_gamma_0}-\ref{fig:evol_gauss_gamma_2} are normalized to the nucleation temperature $T_n$. However, for the benchmark points, we adopt specific numerical values of $T_n$ to illustrate the corresponding numerical values of the initial PBH masses. We take $T_n=100~\mathrm{GeV}$ as a representative value close to the electroweak scale. For the benchmark choices $\alpha=2$ and $\beta/H_n=20$, we obtain $t_n=1.300\times t_{eq}$. The minimum delay required for PBH formation is $t_{PBH}=2.230\times t_{eq}$, which corresponds to an initial PBH mass of $7.09\times10^{-7}M_\odot$ for $T_n=100~\mathrm{GeV}$ (see BP~1 in Table~\ref{tab:BP}).

We focus on moderately strong FOPTs with $\alpha=\mathcal{O}(1)$, as very small values of $\alpha$ generally lead to a suppressed GW signal~\cite{Hindmarsh:2015qta, Hindmarsh:2013xza}. Since we also consider the complementary GW signatures of the phase transition, a sufficiently large $\alpha$ is desirable to obtain a signal with potentially detectable amplitude in current or near-future GW experiments. On the other hand, taking $\alpha\gg1$ can raise concerns regarding the completion and percolation of the phase transition~\cite{Ellis:2018mja}. We therefore restrict our analysis to moderately large values of $\alpha$.

A similar consideration applies to $\beta/H_n$ as well. A very large $\beta/H_n$ corresponds to a rapid phase transition, leaving less time for the delayed phase to develop, while an excessively small $\beta/H_n$ corresponds to a prolonged transition and may raise concerns about the completion of the transition and percolation~\cite{Costa:2025csj}. We therefore choose $\beta/H_n=20$ as a moderate value that allows for a sufficiently prolonged transition and the development of the delayed phase while maintaining a viable completion of the phase transition.

The initial mass of PBH has been shown to depend heavily on the formation temperature ($T_{in}$) in Eq.~\eqref{eq:BHmass}. Furthermore, the formation temperature is dependent upon the collapse dynamics, particularly on the friction and the nucleation rate in this case. Hence, the wall friction and the bubble nucleation rate can have important consequences on the initial mass of PBH.

In what follows, we will explore the consequences of friction and then move on to understand the effects of a Gaussian nucleation rate in comparison to the exponentially decaying rate for PBH formation.

\subsection{Effect of friction}
\label{subsec:fric}

During the cosmological FOPT, the nucleated bubbles expand as the transition from the false vacuum to the true vacuum takes place in the regions surrounding the bubbles. The bubble wall velocity ($v_w$) associated with such expansion can reach relativistic limits almost instantaneously. This can be seen in Fig.~\ref{fig:gamma} of appendix \ref{sec:gamma}, where the lorentz factor corresponding to the velocity of bubble walls are plotted. However, the bubble walls also experience the force of friction that acts against the expansion. This frictional force can arise from several different physical mechanisms, whose contributions to the total friction can vary significantly depending on their origin. Based on the strength of the frictional force, the wall velocities could either go through a run away scenario and over time reach ultra-relativistic ranges, or could be regulated to reach a terminal velocity\footnote{Note that the bubble collision might occur before the $v_w$ reaches the terminal value. In which case, the $\gamma_w (=1/\sqrt{1-v_w^2})$ at the time of collision can be approximately proportional to $T_n/\beta$ \cite{Gouttenoire:2021kjv}. For our the benchmark choices the terminal value is reached well before the bubble collisions as discussed in Appendix~\ref{sec:gamma_c} and hence, we have ignored this possibility in our analysis.}. Hence, it is of utmost importance to understand the different sources of friction and there effects on the expansion of bubble walls.

As the bubbles expand and the vacuum transitions to the broken phase from the unbroken phase, the particles become massive. There is an energy cost associated with this process which can be the leading contributor to the friction. The change in the momentum of such particles can be given by $\Delta p=E-\sqrt{E^2-(\Delta m)^2}$,  where $\Delta m$ is the change of mass and $E$ is the energy of the particle. Since the particles are massless in the initial unbroken phase, $\Delta m$ is simply equal to their masses in the true vacuum. In the high-temperature limit, it can be approximated as $\Delta p\approx\frac{(\Delta m)^2}{2E}$ and the corresponding friction term can be calculated as \cite{Bodeker:2009qy}
\begin{equation}
P_{fric}=\sum_ag_a\int_0^\infty\frac{d^3p}{(2\pi)^3}\frac{(\Delta m_a)^2/(2p)}{\exp(p/T_n)\pm1}
=\sum_ag_ac_a\frac{(\Delta m_a)^2T_n^2}{24},
\label{eq:fric}
\end{equation}
where $g_a$ is the internal number of degrees of freedom of particle $a$. Since Eq.~\eqref{eq:fric} is derived under the high-temperature approximation, the sum runs only over particle species whose masses satisfy $\Delta m_a\ll T_n$. The $c_a$ takes values of 1 and 1/2 for bosons and fermions respectively in accordance with the statistics they follow. This leading order contribution is derived without taking into account the interaction between the neighbouring particles which will introduce further friction of sub-leading order. Refs.~\cite{Gouttenoire:2021kjv,Bodeker:2017cim,BarrosoMancha:2020fay,Long:2024sqg,Hoche:2020ysm} provide a detailed account of the origin and estimation of friction terms order by order.

With the introduction of the friction, the Eq.~\eqref{eq:evol} needs to be updated as
\begin{equation}
\frac{d\rho_R}{dt}+4H\rho_R=-\frac{d\rho_V}{dt}\bigg(1-\frac{2P_{fric}}{\Delta V}\bigg),
\label{eq:evol_fric}
\end{equation}
where the factor of 2 arises due to the effect on the energy densities of the bubble wall as well as the surrounding plasma \cite{Gouttenoire:2023naa}. During the conversion of latent heat into radiation energy density, a substantial fraction can be dissipated through friction when $P_{\rm fric}$ becomes comparable to $\Delta V/2$, as can be inferred from Eq.~\eqref{eq:evol_fric}. As a result the increment in the radiation energy density would be comparatively slower and less overall. Hence, it would take more time to produce the overdensity that is substantial enough to gravitationally collapse and form PBH i.e. higher values of $t_{PBH}$ would have been required to produce $\delta_c$ compared to the frictionless scenario. Thus, the primordial black holes will form comparatively later and with higher initial masses. 

However, in order for the friction to be significant enough to ascertain the change in the initial PBH mass compared to the frictionless scenario the condition $2P_{fric}\lesssim\Delta V$ has to be achieved as has been indicated earlier. This condition can be re-written using Eqs.~\eqref{eq:fric}, \eqref{eq:delv} and \eqref{eq:alpha} as $\bigg(\frac{2\sum_ag_ac_a(\Delta m_a)^2T_n^2}{24}\bigg)\lesssim \bigg(\frac{\pi^2g_\star(T_{eq})\alpha T_n^4}{30}\bigg)$. Even for the best case scenario where all the particles in the spectrum contribute to this friction term (i.e. $\sum_ag_a\approx g_\star(T_{eq})$), this reduces roughly to $\frac{1}{\alpha}(\frac{\Delta m_a}{T_n})^2\lesssim4~(8)$ for $c_a=1~(1/2)$. This is a very difficult condition to satisfy in the high temperature approximation for which $\Delta m_a\ll T_n$ and for a strong FOPT, we require $\alpha>1$. Hence, one may conclude that the friction contribution from particles acquiring a mass in the broken phase, will always remain negligible (i.e. $2P_{fric}{}\ll \Delta V$) in the dynamics of the relevant energy densities for any generic BSM scenario.

The next set of contributions to friction can arise in a scenario where a particle disintegrates into two (or more) particles while entering the broken phase. The scenario in which a particle radiates a gauge boson \cite{Hoche:2020ysm,Gouttenoire:2021kjv} while entering the broken phase is of particular interest. The rate of such a process is dependent heavily upon the lorentz factor $\gamma_w=1/\sqrt{1-v_w^2}$. This warrants an investigation of the consequences of such high velocities (and thus high $\gamma_w$) that the corresponding friction not only dominates the leading order friction mentioned in Eq.~\eqref{eq:fric}, but also significantly delays the collapse. The exact estimation of $\gamma_w$ can be found by solving the following equation
\begin{equation}
\frac{d\gamma_w}{dt}+3Hv_w^2\gamma_w+\frac{2}{R}v_w\gamma_w=\frac{\Delta V-2P_{fric}}{\sigma}v_w,
\label{eq:gamma_w}
\end{equation}
where $\sigma$ is the wall energy per unit area and scales as $\sim T_n^3$. We have chosen the dimensionless proportionality constant to be an $\mathcal{O}(1)$ number. However, for a larger (smaller) value of the constant, the $\gamma_w$ will take longer (shorter) to reach the same terminal value, which can have important consequences on PBH formation in the scenarios with friction. In what follows, we will ignore the friction contribution from the change of mass (i.e. Eq.~\eqref{eq:fric}) and only consider the contribution from the scenario when the particle radiates a gauge boson. Thus, the quantity $P_{fric}$ in Eqs.~\eqref{eq:evol_fric} and \eqref{eq:gamma_w} are dependent on $\gamma_w$ and the differential Eq.~\eqref{eq:gamma_w} has to be added to the bunch of coupled differential equations that are solved numerically to trace the dynamics of the evolution of relevant energy densities.

\begin{figure}[htb!]
\centering
\includegraphics[scale=0.5]{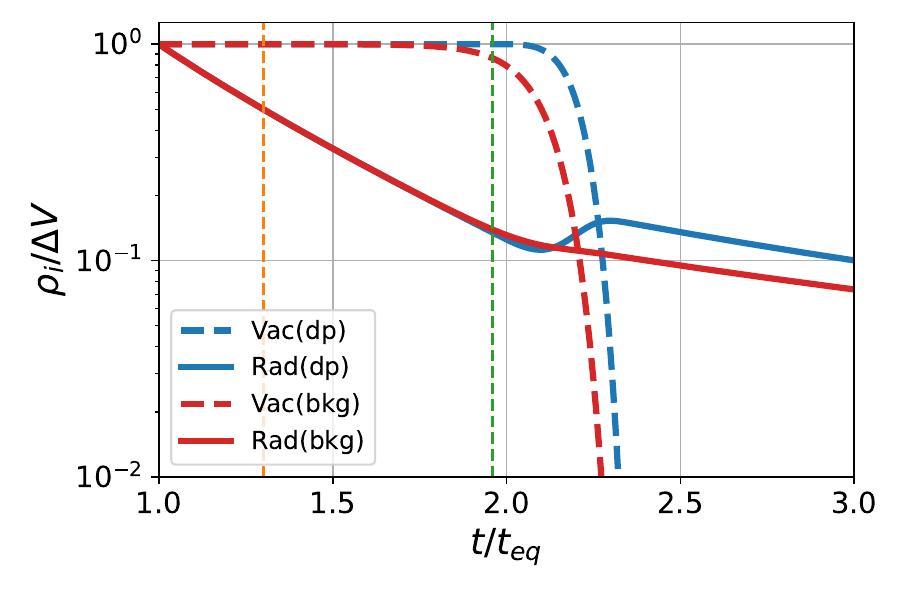}
\includegraphics[scale=0.5]{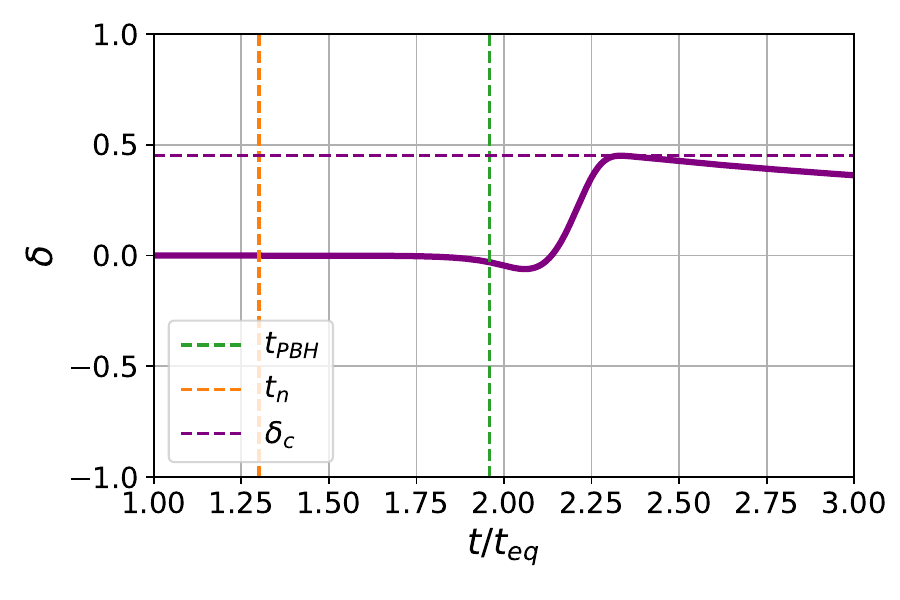}
\caption{\textit{Left:} The evolution of energy densities of vacuum and radiation for background as well as the delayed phase with time for an exponential nucleation rate (Eq.~\eqref{eq:gamma}) with linear friction (Eq.~\eqref{eq:fric1}). The benchmark values used are same as Fig.~\ref{fig:evol_exp_gamma_0}. The minimum delay required to produce PBH is given by $t_{PBH}=1.959\times t_{eq}$. \textit{Right:} Evolution of the $\delta$ parameter that determines PBH formation. This corresponds to BP 2 on Table~\ref{tab:BP}.}
\label{fig:evol_exp_gamma_1}
\end{figure}

When it comes to the calculations of the term $P_{fric}(\gamma_w)$, two prevalent expressions can be found in the literature: one is proportional to $\gamma_w$ \cite{Gouttenoire:2021kjv} while the other is proportional to $\gamma_w^2$ \cite{Hoche:2020ysm}. We will showcase our analysis for both the cases. The friction term that is proportional to the $\gamma_w$ linearly, is given by 
\begin{equation}
P_{fric}(\gamma_w)\sim\frac{\kappa\zeta(3)}{\pi^3}\bigg[\sum_{a,b,c}\nu_ag_aC_{abc}\bigg]\alpha^\prime m_cT_n^3\ln{\bigg(\frac{m_c}{\alpha^{\prime1/2}T_n}\bigg)}\gamma_w,
\label{eq:fric1}
\end{equation}
where $\kappa(\approx4)$ is a dimensionless parameter associated with the momentum transfer arising from interactions between the bubble wall and the particles in the plasma \cite{Gouttenoire:2021kjv}, $\alpha^\prime\approx1/30$, $\nu_a=1(3/4)$ for bosons(fermions), $g_a$ is the number of degrees of freedom of species $a$, $C_{abc}$ is the charge factor and $m_c$ (assumed to be 100 GeV as a benchmark) is the mass of the radiated gauge boson. $\bigg[\sum_{a,b,c}\nu_ag_aC_{abc}\bigg]$ is assumed to be $\sim100$ for the sake of simplicity following Ref.~\cite{Gouttenoire:2021kjv}. Now that an explicit $T_n$ dependence has appeared through the friction term, it is not feasible anymore to draw a frictional counterpart of Fig.~\ref{fig:evol_exp_gamma_0} without assuming a benchmark value. For benchmark values $\alpha=2$, $\beta/H_n=20$ and $T_n=100$ GeV, the time evolution of the relevant energy densities and $\delta$ are shown in Fig.~\ref{fig:evol_exp_gamma_1} where the transfer of vacuum energy to radiation energy is not as efficient anymore due to friction. Hence, despite the reduction in $t_{PBH}$ to 1.959$\times t_{eq}$, the radiation energy densities are much smaller compared to the frictionless scenario, which leads to higher initial masses of PBH (see Eq.~\eqref{eq:BHmass}). The relevant quantities are mentioned in BP 2 of Table~\ref{tab:BP}, where one can see a roughly 147\% rise in the PBH initial mass from BP 1.

\begin{figure}[htb!]
\centering
\includegraphics[scale=0.5]{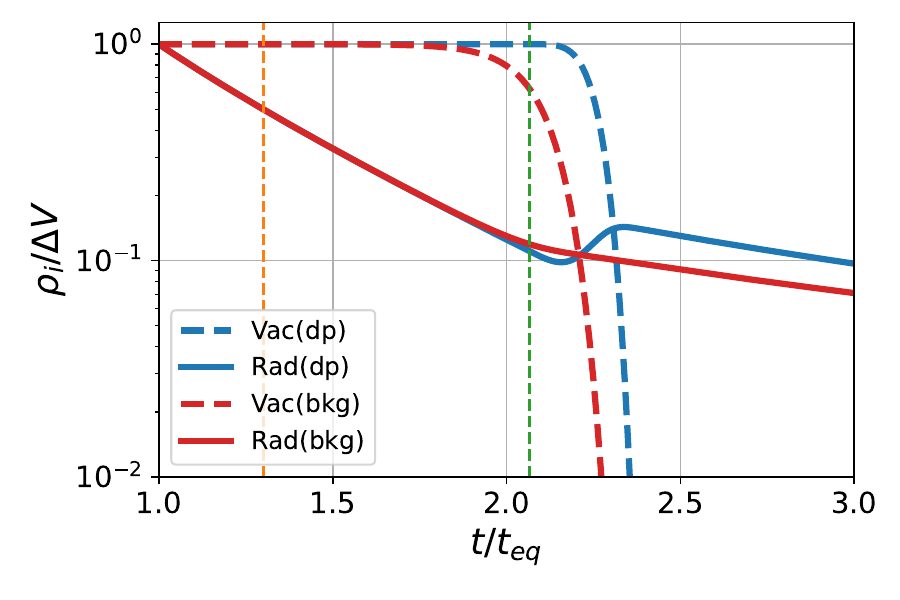}
\includegraphics[scale=0.5]{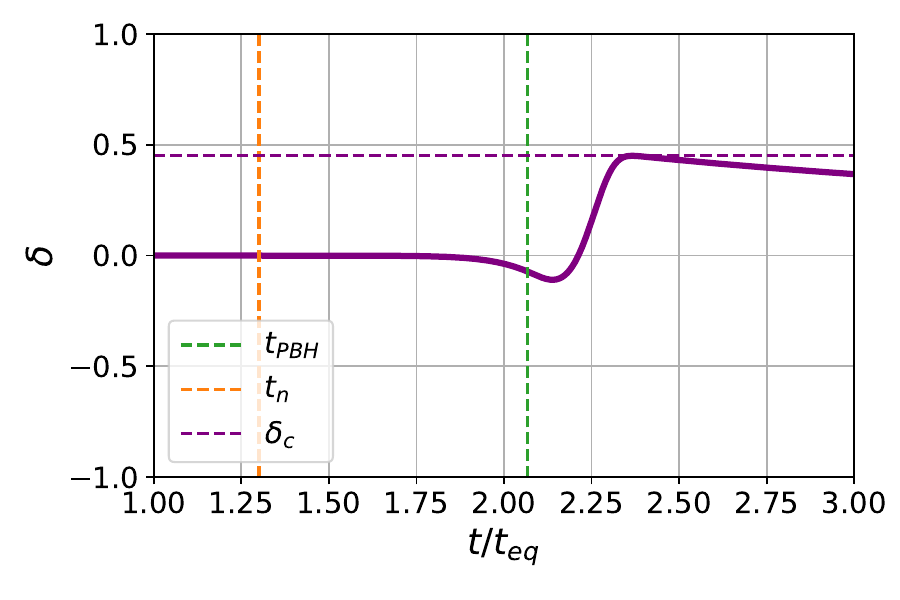}
\caption{\textit{Left:} The evolution of energy densities of vacuum and radiation for background as well as the delayed phase with time for an exponential nucleation rate (Eq.~\eqref{eq:gamma}) with quadratic friction (Eq.~\eqref{eq:fric2}). The benchmark values used are same as Fig.~\ref{fig:evol_exp_gamma_0}. The minimum delay required to produce PBH is given by $t_{PBH}=2.066\times t_{eq}$. \textit{Right:} Evolution of the $\delta$ parameter that determines PBH formation. This corresponds to BP 3 on Table~\ref{tab:BP}.}
\label{fig:evol_exp_gamma_2}
\end{figure}

One can however, choose to follow Ref.~\cite{Hoche:2020ysm} instead and write the next to leading order term in friction as
\begin{equation}
P_{fric}(\gamma_w)\sim\frac{\zeta(3)}{\pi^3}\bigg[\sum_{a,b,c}\nu_ag_aC_{abc}\bigg](2\ln{2}-1)\alpha^\prime T_n^4\gamma_w^2
\label{eq:fric2}
\end{equation}
where the benchmark values related to the parameters remain the same as in the previous case and we obtain Fig.~\ref{fig:evol_exp_gamma_2}. The corresponding benchmark values and relevant quantities are mentioned in BP 3 of Table~\ref{tab:BP} where we can see a 155\% rise in the PBH initial mass compared to BP 1.

In both the cases of friction (Fig.~\ref{fig:evol_exp_gamma_1} and Fig.~\ref{fig:evol_exp_gamma_2}), apart from the less dilution of radiation energy density in the delayed patch compared to the background, there is another factor at play. The bubble nucleation happens late in the delayed patch and hence, the $\gamma_w$ reaches the terminal value later, which results in less friction in the delayed phase just after $t_{PBH}$ compared to the background. This phenomena can be somewhat guessed by looking at the evolution of radiation energy densities in both Fig.~\ref{fig:evol_exp_gamma_1} and Fig.~\ref{fig:evol_exp_gamma_2} (also in Fig.~\ref{fig:evol_gauss_gamma_1} and Fig.~\ref{fig:evol_gauss_gamma_2}). A more explicit verification is given in Appendix \ref{sec:gamma}, where the evolution of the $\gamma_w$'s are shown.

We conclude this section with a discussion on the PBH mass at formation. We have seen that with the introduction of friction, there is a reduction in $t_{PBH}$ compared to the frictionless scenario. This leads to the PBH forming comparatively earlier in the cosmological history. One could naively expect the initial mass of PBH to be smaller since at early times the radiation energy density of the patch is comparatively less diluted due to less red-shifting. However, one also has to take the difference in the energy budget compared to the frictionless case into account. With the addition of friction, a huge chunk of the vacuum energy density is lost to friction and hence, the radiation energy density receives only a very small contribution, which leads to larger initial masses of PBH as is evident in Fig.~\ref{fig:evol_exp_gamma_1} and Fig.~\ref{fig:evol_exp_gamma_2}. This phenomena may have important consequences in the cosmological observations related to the history of the early universe. Now that we have analyzed the effect of friction within regime of the exponential nucleation rate, we can move on to understand the effect of different nucleation rate profiles in Sec.~\ref{subsec:rate}.

\subsection{Effect of bubble nucleation rate}
\label{subsec:rate}

Apart from the introduction of the friction term, the other quantity that can have a significant influence on the production of PBH is the profile of the bubble nucleation rate. The rate of bubble nucleation can increase exponentially after the nucleation temperature as has been assumed in Eq.~\eqref{eq:gamma}. However, in certain cases, especially where the FOPT is substantially supercooled, the temperature dependence of the three-dimensional Euclidean action (often represented as $S_3$) could be such that the rate is maximized at a temperature around nucleation\footnote{Further subtleties can be introduced by considering the maxima to be substantially away from nucleation \cite{Goncalves:2024vkj} which is avoided here for simplification.} and is suppressed again over time \cite{Kanemura:2024pae,Cutting:2018tjt,Costa:2025csj,Goncalves:2024vkj,Lewicki:2024sfw}. This phenomena can be modeled with a nucleation rate given by
\begin{equation}
\tilde\Gamma_V (t)=H(T_n)^4e^{-\frac{1}{2}\tilde\beta^2(t-t_n)^2},
\label{eq:gammat}
\end{equation}
where $\tilde\beta^{-1}$ is again a measure of the duration when the phase transition is most effective. While the temperature dependence of the parameter $S_3/T$ is largely a model dependent phenomena, the consequences that we are interested in can nevertheless be studied in a generic manner. In what follows, the effect of different nucleation rates (given by Eqs.~\eqref{eq:gamma} \& \eqref{eq:gammat}) on the dynamics of delayed phase transition is studied to delineate the change in the initial mass of the PBH that is formed due to the collapse of over-densities.

\begin{figure}[htb!]
\centering
\includegraphics[scale=0.5]{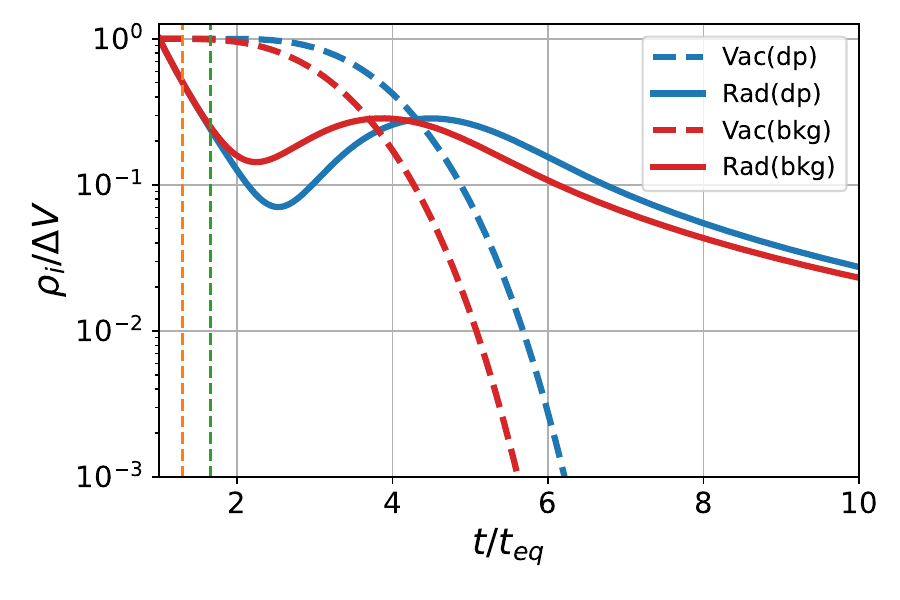}
\includegraphics[scale=0.5]{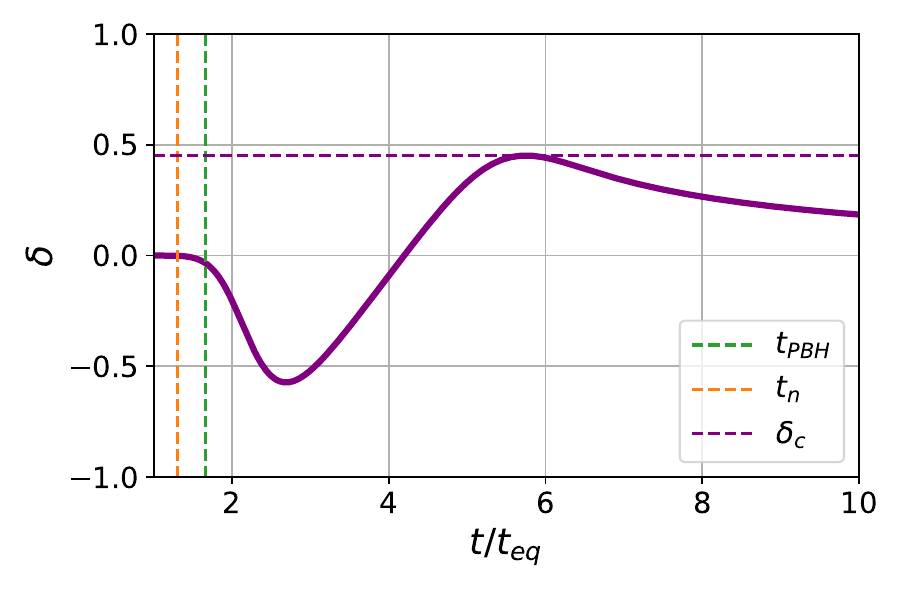}
\caption{\textit{Left:} The evolution of energy densities of vacuum and radiation for background as well as the delayed phase with time for a Gaussian nucleation rate (Eq.~\eqref{eq:gammat}) without taking into account the effect of friction. The benchmark values used are $\alpha=2$ and $\beta/H_n=20$ which provides $t_n=1.300\times t_{eq}$. The minimum delay required to produce PBH is given by $t_{PBH}=1.660\times t_{eq}$. \textit{Right:} Evolution of the $\delta$ parameter that determines PBH formation. This corresponds to BP 4 on Table~\ref{tab:BP}.}
\label{fig:evol_gauss_gamma_0}
\end{figure}

\begin{figure}[htb!]
\centering
\includegraphics[scale=0.5]{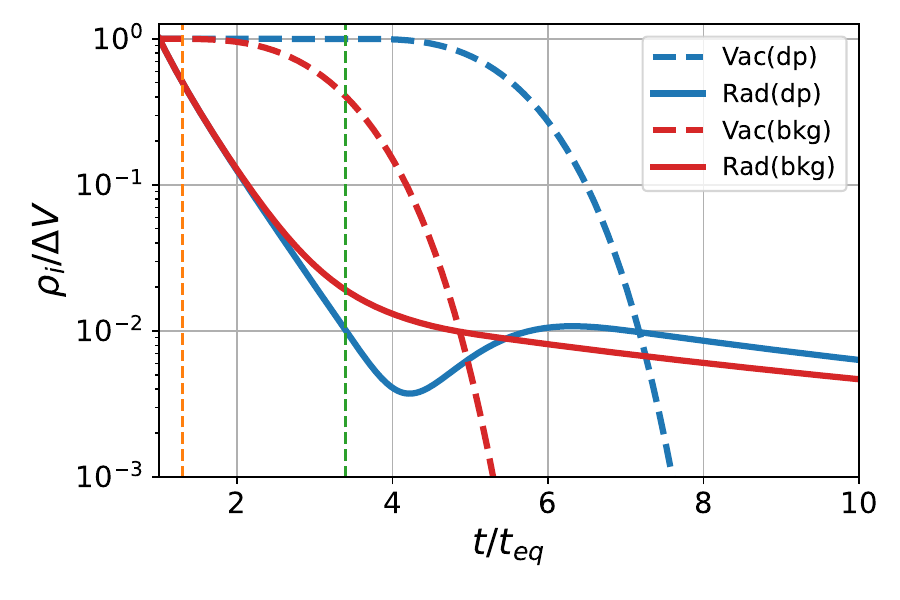}
\includegraphics[scale=0.5]{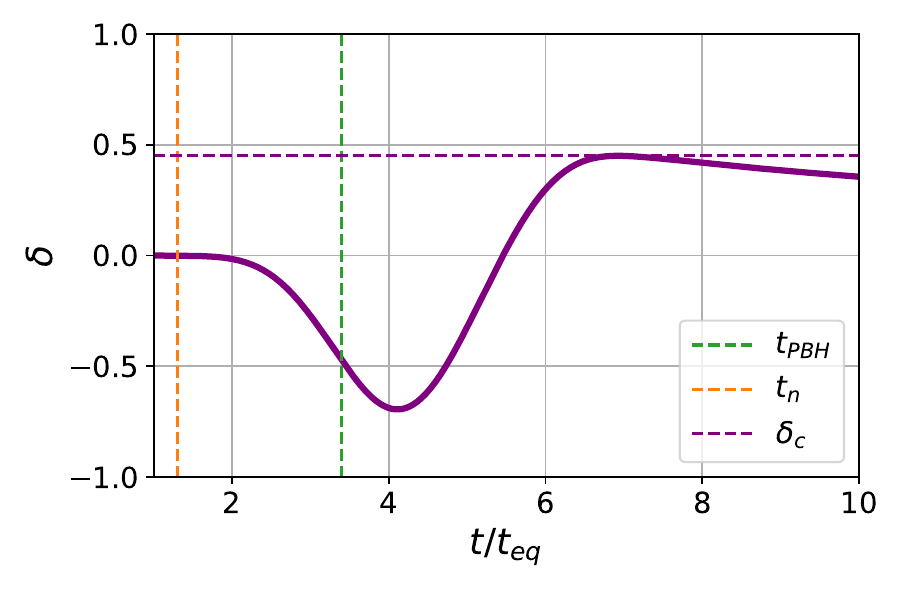}
\caption{\textit{Left:} The evolution of energy densities of vacuum and radiation for background as well as the delayed phase with time for a Gaussian nucleation rate (Eq.~\eqref{eq:gammat}) with linear friction (Eq.~\eqref{eq:fric1}). The benchmark values used are same as Fig.~\ref{fig:evol_gauss_gamma_0}. The minimum delay required to produce PBH is given by $t_{PBH}=3.391\times t_{eq}$. \textit{Right:} Evolution of the $\delta$ parameter that determines PBH formation. This corresponds to BP 5 on Table~\ref{tab:BP}.}
\label{fig:evol_gauss_gamma_1}
\end{figure}

\begin{figure}[htb!]
\centering
\includegraphics[scale=0.5]{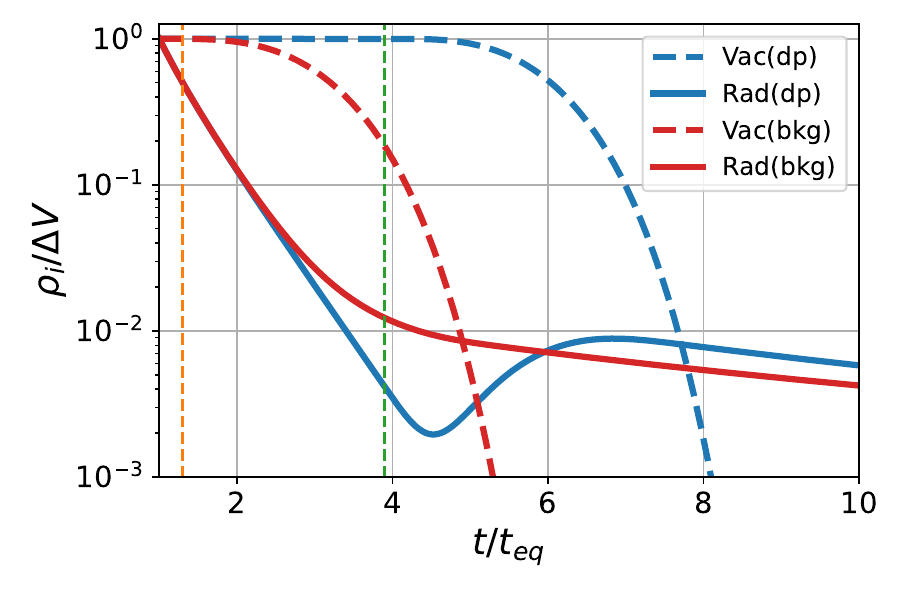}
\includegraphics[scale=0.5]{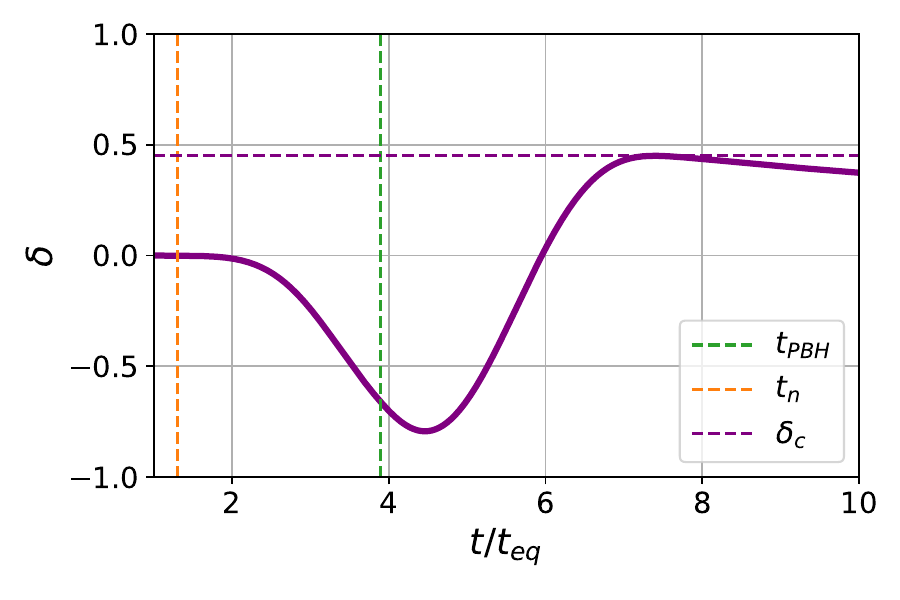}
\caption{\textit{Left:} The evolution of energy densities of vacuum and radiation for background as well as the delayed phase with time for a Gaussian nucleation rate (Eq.~\eqref{eq:gammat}) with quadratic friction (Eq.~\eqref{eq:fric2}). The benchmark values used are same as Fig.~\ref{fig:evol_gauss_gamma_0}. The minimum delay required to produce PBH is given by $t_{PBH}=3.892\times t_{eq}$. \textit{Right:} Evolution of the $\delta$ parameter that determines PBH formation. This corresponds to BP 6 on Table~\ref{tab:BP}.}
\label{fig:evol_gauss_gamma_2}
\end{figure}

In order to compare the effects of different nucleation rates, the quantities $\beta$ and $\tilde\beta$ would have to be treated on the equal footing. A connection between these quantities can be drawn through bubble number density ($n_b$) which is related to the $\beta$ and $\tilde\beta$ as
\begin{equation}
n_b=\frac{\beta^3}{8\pi}~~~~\textrm{and}~~~~ n_b=\frac{\sqrt{2\pi}H_n^4}{\tilde{\beta}},
\label{eq:num_den}
\end{equation}
where, the wall velocity is assumed to be of order unity in natural units \cite{Cutting:2018tjt}. To effectively draw comparisons between the exponential and Gaussian rates, Eq.~\eqref{eq:num_den} can be rewritten as the following
\begin{equation}
\tilde{\beta}=\frac{\sqrt{2\pi}H_n^4}{\beta^3}8\pi.
\label{eq:betat}
\end{equation}
The same value of $\beta$, as used in Fig.~\ref{fig:evol_exp_gamma_0}, is chosen and the corresponding $\tilde\beta$ is found using Eq.~\eqref{eq:betat}, which is used in Eq.~\eqref{eq:gammat} to study the evolution of energy densities to arrive at Fig.~\ref{fig:evol_gauss_gamma_0}. The $t_{PBH}$ required to produce the critical overdensity decreases from 2.23 in BP 1 to 1.66 in $t_{eq}$ units when a Gaussian nucleation profile is chosen instead of exponential. Despite the reduction in $t_{PBH}$, it takes longer for the overdensity to reach the critical value due to the modified nucleation rate. The contribution to the radiation energy densities are also reduced in Fig.~\ref{fig:evol_gauss_gamma_0} as compared to Fig.~\ref{fig:evol_exp_gamma_0} which leads to the formation of PBHs with greater initial mass. The details of a PBH formation from delayed FOPT for a Gaussian nucleation rate in the frictionless scenario are given in BP 4 of Tab.~\ref{tab:BP} where a 124\% rise in the PBH initial mass is evident.


\begin{table}[htb!]
\centering
\setlength{\tabcolsep}{7pt}
\renewcommand{\arraystretch}{1.3}

\begin{tabular}{|c|c|c|c|c|c|c|c|c|}
\hline
BPs & $\alpha$ & $\beta/H_n$ & $T_n$ (GeV) & Nucleation Rate &$P_{\rm fric}$ & $t_{PBH}$ ($t_{eq}$)& $M_{BH}^{\rm in}$ ($M_\odot$) & Figure \\ \hline\hline
BP 1 & 2 & 20 & $10^2$ & exponential & 0 & 2.230 & $7.09\times10^{-7}$ & Fig.~\ref{fig:evol_exp_gamma_0}\\ \hline
BP 2 & 2 & 20 & $10^2$ & exponential & $\sim\gamma_w$ & 1.959 & $1.75\times10^{-6}$ &Fig.~\ref{fig:evol_exp_gamma_1}\\ \hline
BP 3 & 2 & 20 & $10^2$ & exponential & $\sim\gamma_w^2$ & 2.066 & $1.81\times10^{-6}$&Fig.~\ref{fig:evol_exp_gamma_2} \\ \hline\hline
BP 4 & 2 & 20 & $10^2$ & Gaussian & 0 & 1.660 & $1.58\times10^{-6}$ &Fig.~\ref{fig:evol_gauss_gamma_0}\\ \hline
BP 5 & 2 & 20 & $10^2$ & Gaussian & $\sim\gamma_w$& 3.391  & $6.73\times10^{-6}$&Fig.~\ref{fig:evol_gauss_gamma_1} \\ \hline
BP 6 & 2 & 20 & $10^2$ & Gaussian & $\sim\gamma_w^2$ & 3.892 & $7.39\times10^{-6}$&Fig.~\ref{fig:evol_gauss_gamma_2} \\ \hline
\end{tabular}

\caption{Benchmark values describing the delayed phase transition and the formation of PBH.}
\label{tab:BP}
\end{table}

In this section so far, the Gaussian nucleation profile is compared with the exponential scenario within the frictionless assumption. In what follows, we introduce friction term linear as well as quadratic in $\gamma_w$ within the Gaussian nucleation paradigm and perform the analysis. We found that the nucleation in the delayed patch (i.e. $t_{PBH}$) has to be significantly delayed in both cases of friction to achieve the critical overdensity to form PBH as mentioned in BP 5 and 6 of Table~\ref{tab:BP}. Furthermore, with the addition of friction, the transfer of vacuum energy density to the radiation energy density is very significantly affected as is depicted in Fig.~\ref{fig:evol_gauss_gamma_1} and Fig.~\ref{fig:evol_gauss_gamma_2} and the initial masses of PBHs see a rise of over 300\% in both cases (BP 5 \& 6) compared to the frictionless scenario (BP 4). Finally, PBH production from delayed phase transition can conjure constraints from over-closure of the Universe \cite{Liu:2021svg} and dark matter overdensity (from micro-lensing) \cite{Hashino:2021qoq} depending on the time taken for the phase transition. The benchmark choice of $\beta/H_n$ made in this work, however, is rather large and hence it does not succumb to such constraints. 

Our next aim is to shed complementary light on the PBH mass spectrum for different benchmark scenarios using their corresponding GW signatures. This is motivated by the fact that a strong first-order phase transition (FOPT) in the early Universe can potentially generate detectable GW signals.

Over the past decade, GW astronomy has advanced remarkably with observations from pulsar timing arrays (PTAs) in the nano-Hertz band~\cite{NANOGrav:2023gor,NANOGrav:2023hde,NANOGrav:2023hvm,Reardon:2023gzh,EPTA:2023fyk,Xu:2023wog,Antoniadis:2022pcn} and the LIGO-Virgo-KAGRA (LVK) collaboration~\cite{LIGOScientific:2016wof} in the kilo-Hertz range. This rapid progress has opened a new window to probe physics beyond the Standard Model through GW signals spanning a broad frequency spectrum, from nano- to kilo-Hertz.

Future GW experiments will further extend this coverage. In the nano-Hertz regime, PTAs such as SKA~\cite{Janssen:2014dka} will play a crucial role. In the milli-Hertz band, space based interferometers like LISA~\cite{LISA:2017pwj}, $\mu$Ares~\cite{Sesana:2019vho}, AEDGE~\cite{AEDGE:2019nxb}, and atom interferometer based proposals such as AION~\cite{Badurina:2019hst} are expected to provide enhanced sensitivity. In the deci-Hertz regime planned missions such as DECIGO~\cite{Kawamura:2020pcg}, UDECIGO~\cite{Kudoh:2005as} and BBO~\cite{Harry:2006fi} will bridge the gap between space and ground based detectors. At higher frequencies next generation ground based detectors including the Einstein Telescope (ET)~\cite{Hild:2008ng} and Cosmic Explorer (CE)~\cite{LIGOScientific:2016wof} will significantly improve sensitivity in the kilo-Hertz range.

In the next section, we first discuss about the GW spectra arising from FOPTs in the early Universe for different scenarios: `with friction' and `without friction' (considering both exponential and Gaussian nucleation profiles) and examine whether these scenarios can lead to distinctive GW imprints within the detectable frequency range.


\section{Gravitational wave emission from First-Order Phase Transition}
\label{sec:GW}

During a FOPT, the expansion and interaction of bubbles of the true vacuum with the surrounding plasma can generate GWs through three primary mechanisms: bubble wall collisions, the generation of sound waves in the plasma, and magnetohydrodynamic turbulence.

In absence of friction, $\gamma_w$ keeps increasing with time from bubble formation until collision (see the top left and middle right panels of Fig.~\ref{fig:gamma} for example), resulting in runaway walls and the walls never reach a terminal velocity for either exponential or Gaussian nucleation profiles. In this case, most of the vacuum energy is transferred into the bubble wall motion, while the plasma is not efficiently excited. As a result, the GW signal is predominantly sourced by the bubble collision contribution~\cite{Gouttenoire:2021kjv}. However, for the benchmark value $\alpha = 2$ chosen for illustration (see Table~\ref{tab:BP} for details) we find that the amplitude of the GW signal is very weak and lies well below the future detector sensitivity.

On the other hand, if a next-to-leading order (NLO) $\gamma_w$-dependent friction is present in the theory then the scenario can become completely different.\footnote{For our analysis we consider that the leading order $\gamma_w$ independent friction is negligible; therefore, by `with-friction' we mean the NLO $\gamma_w$-dependent friction.} On top of bubble collisions, sound waves and magnetohydrodynamic (MHD) turbulence can be generated in the plasma due to the transfer of vacuum energy into the fluid, which can serve as the primary GW sources~\cite{Gouttenoire:2021kjv}. If the terminal value of $\gamma_w$, $\gamma_t$ is grater than the value of $\gamma_w$ at the time of collision, $\gamma_c$, the conclusion remains similar to the runaway case discussed above~\cite{Gouttenoire:2021kjv}. 

However, for all `with-friction' benchmark points we find that the bubble wall velocity saturates almost immediately after formation, as shown by the red curves in Fig.~\ref{fig:gamma}. Since the GW spectra are primarily controlled by the evolution of the dominant background patches and for our benchmarks we find $\gamma_c \gg \gamma_t$; therefore, one finds that a strong fluid motion can develop and most of the vacuum energy can be transferred to the plasma, making sound waves and turbulence the dominant sources of GWs. We present the approximate estimation of $\gamma_c$ for `with-friction' benchmark points in the Appendix~\ref{sec:gamma_c}. 

In addition to the FOPT, scalar-induced gravitational waves (SIGWs) generated by curvature perturbations can also contribute to the GW background. However, recent studies have shown that the SIGW spectrum is typically subdominant to the FOPT-induced GW signal~\cite{Franciolini:2025ztf}. In the following paragraphs, we briefly discuss the GW spectra arising from sound waves and MHD turbulence one after another and present their forms as functions of the FOPT parameters $\{\alpha, \beta, T_n\}$.

Following the above discussion, we assume $v_w\simeq1$. This is justified by the fact that in the presence of friction $\gamma_w$ terminates at $\gamma_t\gg1$ (see Fig.~\ref{fig:gamma} for an example). With this approximation, the peak amplitude and the corresponding peak frequency of the sound-wave-driven GW spectrum are estimated as follows:~\cite{Hindmarsh:2015qta, Hindmarsh:2013xza}
\begin{align}
h^2\Omega^{\text{sw}}_{\text{peak}}\simeq2.7\times 10^{-6}\kappa_v^2\Upsilon_{\rm SW}\left(\frac{H_n}{\beta}\right)\left(\frac{\alpha}{1+\alpha}\right)^2\left(\frac{g_*}{100}\right)^{-\frac{1}{3}},
\label{Eq:Omega_sw_peak}
\end{align}
\begin{align}
f_{\text{peak}}^{\text{sw}}\simeq19\left(\frac{\beta}{H_n}\right)\left(\frac{T_n}{10^8 \text{GeV}}\right)\left(\frac{g_*}{100}\right)^{1/6} \text{Hz}.
\label{Eq:f_sw_peak}
\end{align}
Here, the parameter $\Upsilon_{\rm SW}$ quantifies the suppression arising due to the finite life time of the sound waves, which we take to be $\simeq 10^{-3}$ as a benchmark. The other parameter $\kappa_v$ follows the form, $\kappa_v=\alpha/(0.73+0.083\sqrt{\alpha}+\alpha)$~\cite{Espinosa:2010hh}.

The total sound wave contribution in terms of $\Omega^{\text{sw}}_{\text{peak}}$ and $f_{\text{peak}}^{\text{sw}}$ is given by~\cite{Caprini:2015zlo,Jana:2025vyb}
\begin{align}
\Omega^{\text{sw}}_{\text{GW}}(f)\simeq\Omega^{\text{sw}}_{\text{peak}}\left(\frac{7}{4+3(\frac{f}{f_{\text{peak}}^{\text{sw}}})^2}\right)^{7/2}\left(\frac{f}{f_{\text{peak}}^{\text{sw}}}\right)^3.
\label{Eq:Omega_sw}
\end{align}

For the benchmark ${\alpha=2,\ \beta=20,\ T_n=100~\mathrm{GeV}}$ considered here, we find that the overall amplitude of the GW spectrum is predominantly determined by the sound-wave contribution, $\Omega^{\text{sw}}_{\mathrm{GW}}(f)$. However, the high-frequency tail of the spectrum is typically dominated by the contribution from MHD turbulence.

Analogous to the sound wave component, the peak amplitude and peak frequency of the GW spectrum sourced by turbulence can be approximated as~\cite{Kamionkowski:1993fg}
\begin{align}
h^2\Omega^{\text{tur}}_{\text{peak}} \simeq 3.4\times 10^{-4}
\left(\frac{H_n}{\beta}\right)
\left(\frac{\kappa_{\text{tur}}\alpha}{1+\alpha}\right)^{3/2}
\left(\frac{g_*}{100}\right)^{-1/3},
\label{Eq:Omega_tur_peak}
\end{align}
\begin{align}
f_{\text{peak}}^{\text{tur}} \simeq 27
\left(\frac{\beta}{H_n}\right)
\left(\frac{T_n}{10^8\,\text{GeV}}\right)
\left(\frac{g_*}{100}\right)^{1/6} \,\text{Hz},
\label{Eq:f_tur_peak}
\end{align}
where $\kappa_{\text{tur}} \simeq 0.05\,\kappa_v$.

The turbulence contribution to the GW spectrum, expressed in terms of $\Omega^{\text{tur}}_{\text{peak}}$ and $f_{\text{peak}}^{\text{tur}}$, is given by~\cite{Caprini:2009yp,Caprini:2015zlo,Jinno:2015doa}
\begin{align}
\Omega^{\text{tur}}_{\text{GW}}(f) \simeq 
\Omega^{\text{tur}}_{\text{peak}}
\left(\frac{f}{f_{\text{peak}}^{\text{tur}}}\right)^3 
\frac{\left(1+ \frac{8\pi f}{h_{*,n}}\right)^{-1}}
{\left(1+\frac{f}{f_{\text{peak}}^{\text{tur}}}\right)^{11/3}},
\label{Eq:Omega_tur}
\end{align}
where
\begin{align}
h_{*,n} = 17
\left(\frac{g_*}{100}\right)^{1/6}
\left(\frac{T_n}{10^8\,\text{GeV}}\right)\,\text{Hz}.
\label{Eq:h_star}
\end{align}

\begin{figure}
    \centering
    \includegraphics[width=.85\linewidth]{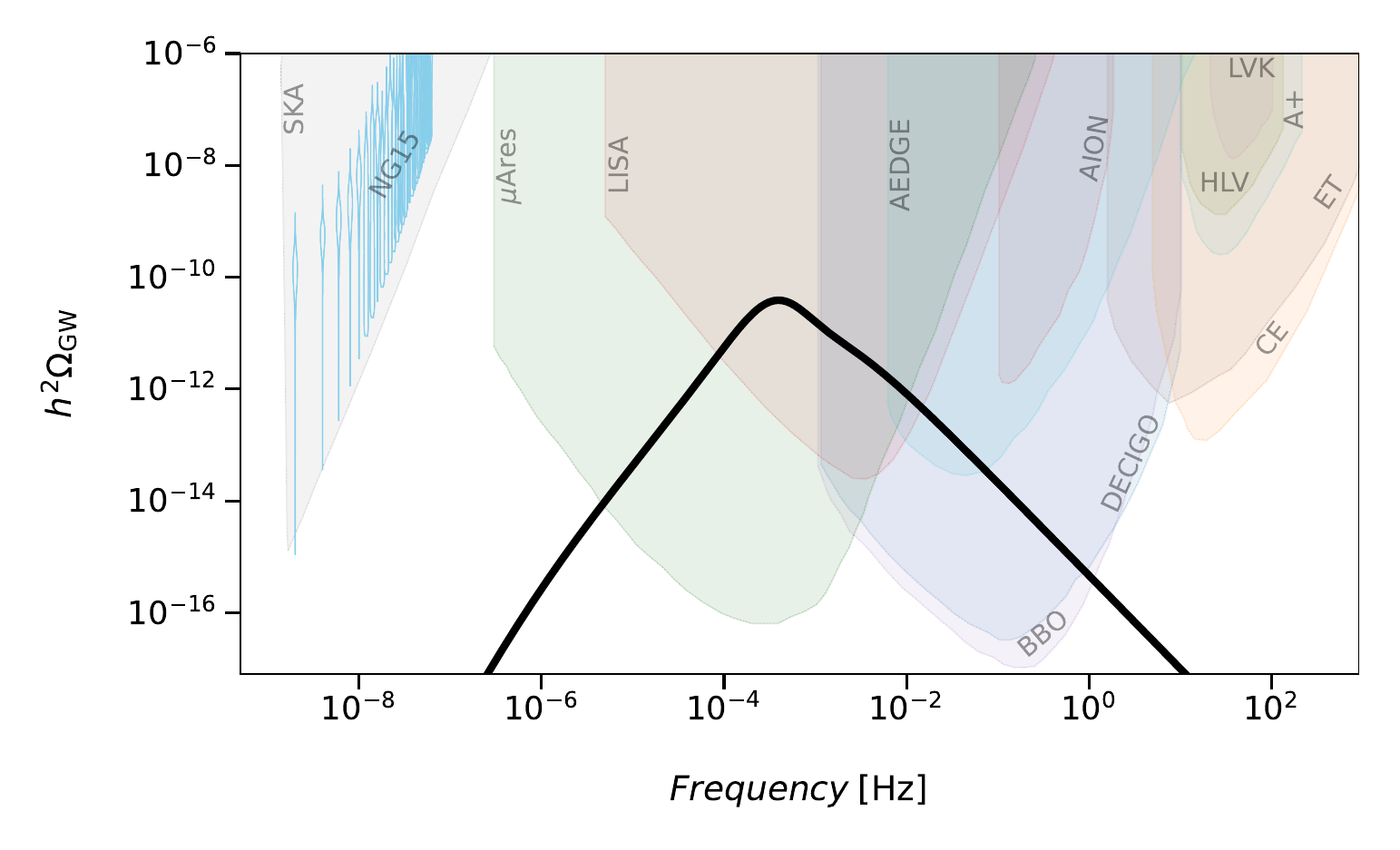}
    \caption{The gravitational wave spectra originated from the FOPT for benchmark values $\alpha=2$, $\beta/H_n=20$ and $T_n=100$ GeV (consistent with BP~2, BP~3, BP~5, and BP~6).}
    \label{fig:gw}
\end{figure}

Using the above expressions for the GW spectra we evaluate the stochastic signal for the benchmark `with-friction' exponential nucleation parameters $\left\{ \alpha = 2, \beta/H_n = 20, T_n = 10^2\,\text{GeV} \right\}$ (BP~2) and present the result in Fig.~\ref{fig:gw} as the black solid line. Performing the same analysis for the Gaussian nucleation parameters $\left\{ \alpha, \tilde{\beta}, T_n \right\}$ (BP~5), while keeping $T_n$ and $\alpha$ fixed and choosing $\tilde{\beta}$ such that it reproduces the same value of $n_b$ as in the exponential case, yields an identical GW spectrum~\cite{Caprini:2015zlo,Costa:2025csj}.

This can be understood from the fact that the parameter $n_b$ is more fundamental, with both $\beta/H_n$ and $\tilde{\beta}$ being related to $n_b$ through Eq.~(\ref{eq:num_den}). Furthermore, the slight difference in the $\gamma_w$ profile arising from NLO friction (e.g., $\gamma_w$ versus $\gamma_w^2$ dependence) does not significantly alter the bubble dynamics, as in both cases $\gamma_w$ saturates shortly after nucleation (see Fig.~\ref{fig:gamma}). As a result, the GW signatures for BP~2, BP~3, BP~5, and BP~6 are indistinguishable, and the black solid line in Fig.~\ref{fig:gw} effectively represents all four benchmarks.

As a consequence, the microphysical origin of a FOPT cannot be uniquely inferred from the GW spectrum alone: different microphysics can lead to the same $\left\{ \alpha, n_b, T_n \right\}$ while predicting different PBH masses, yet producing indistinguishable GW spectra. Therefore, this intrinsic degeneracy in PBH mass estimation cannot be resolved using GW observations alone.

In summary, BP~1 and BP~4 (benchmarks without friction) do not yield a detectable GW signal, whereas  BP~2, BP~3, BP~5, and BP~6 (benchmarks with friction) produce indistinguishable signals, as shown in Fig.~\ref{fig:gw}.

\section{Conclusion}
\label{sec:conc}

First order phase transition in the early universe is a very interesting phenomena from the standpoint of baryogenesis, stochastic GWs, and various BSM scenarios. The detection of GWs originated from FOPT can provide us with important information regarding the vacuum structure of the universe. Furthermore, due to the formation of causally connected delayed patches, overdensities can be created, which under appropriate conditions can gravitationally collapse and form PBHs. We have seen that the friction, introduced due to interactions of the particles as they enter the bubbles, play a very important role in the collapse and formation of PBHs. Furthermore, especially in the supercooled regime, the nature of the nucleation profile (exponential and Gaussian) can induce significant consequences as has been shown in this work. Specifically, the effects can be summed up as the following:
\begin{itemize}
\item In a FOPT within gauge theories, particles entering the bubble can radiate gauge boson(s), leading to a velocity-dependent frictional force on the bubble walls (we restrict our discussion of friction to gauge theories). As a result, once the transition takes place, a significant fraction of the vacuum energy stored in both the delayed patch and the background is dissipated into friction. As a consequence, the radiation energy densities receive a significantly smaller contribution w.r.t. the frictionless scenario. Therefore, when the critical overdensity is reached, PBHs with significantly higher initial masses are produced compared to the frictionless scenario because of the inverse square-root dependence of the PBH initial mass on the radiation energy density.

\item Furthermore, the nucleation rate itself can follow a Gaussian profile instead of the generally assumed exponential profile. We observe that for a Gaussian nucleation rate, the whole phenomena takes longer due to its extended profile. Consequently, even if the nucleation time ($t_{\rm PBH}$) is less delayed in the delayed patch compared to the exponential case, the required overdensity is reached much later. Meanwhile, the radiation energy densities keep redshifting, leading to formation of PBHs with higher initial masses compared to the scenario with exponential nucleation rate.

\end{itemize}
In addition, we argue that the stochastic GW signal from a FOPT, which we try to use as a complementary probe of the PBH mass spectrum, can potentially distinguish between scenarios with and without friction. In one case (without friction scenario), the signal is dominated by bubble collisions and is unlikely to be detectable in the near future due to its low amplitude, while in the other case (with friction scenario), the signal is dominated by sound waves and MHD turbulence and the corresponding spectrum is shown in Fig.~\ref{fig:gw}. However, we observe that the GW signal is unable to distinguish between exponential and Gaussian nucleation profiles if we consider the same bubble number density ($n_b$) at the time of nucleation for both cases.

The location of the stochastic GW peak can provide important information about the nucleation temperature and the $\beta$ (or $\tilde\beta$) parameter, while the details of friction and nucleation rate profile (including $\beta$ or $\tilde\beta$) can be linked to PBH formation. With better theoretical understanding of the precise value of critical overdensity ($\delta_c$) and more realistic PBH initial mass distributions in future, probes of PBH can provide an additional handle to understanding the dynamics of FOPT in the early universe in conjuction with the stochastic GWs.

\acknowledgments
KL wants to thank Joydeep Chakrabortty for useful comments. The authors acknowledge the support of Harish-Chandra Research Institute and Homi Bhabha National Institute (HBNI), Mumbai.

\appendix
\section{Evolution of bubble walls}
\label{sec:gamma}
\begin{figure}[htb!]
\centering
\includegraphics[scale=0.5]{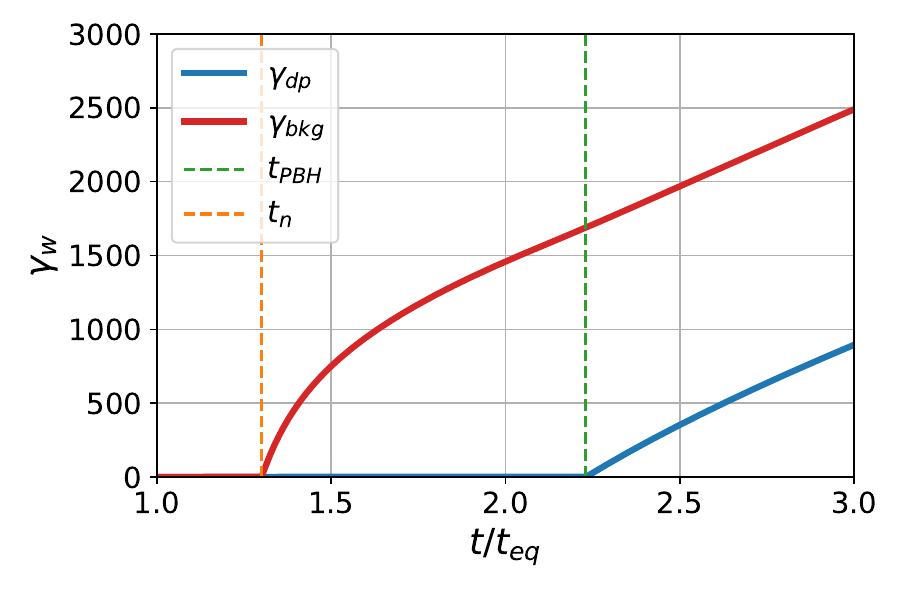}
\includegraphics[scale=0.5]{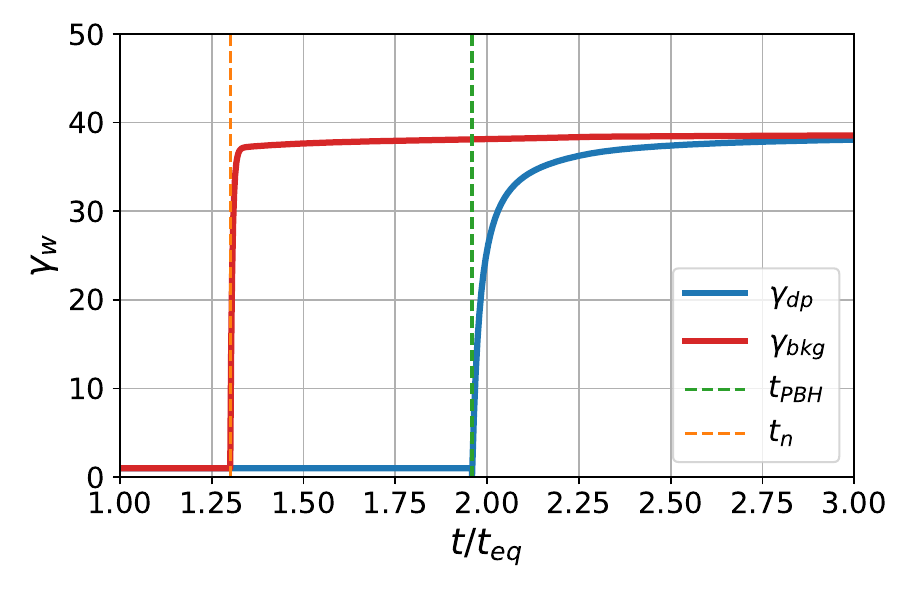}\\
\includegraphics[scale=0.5]{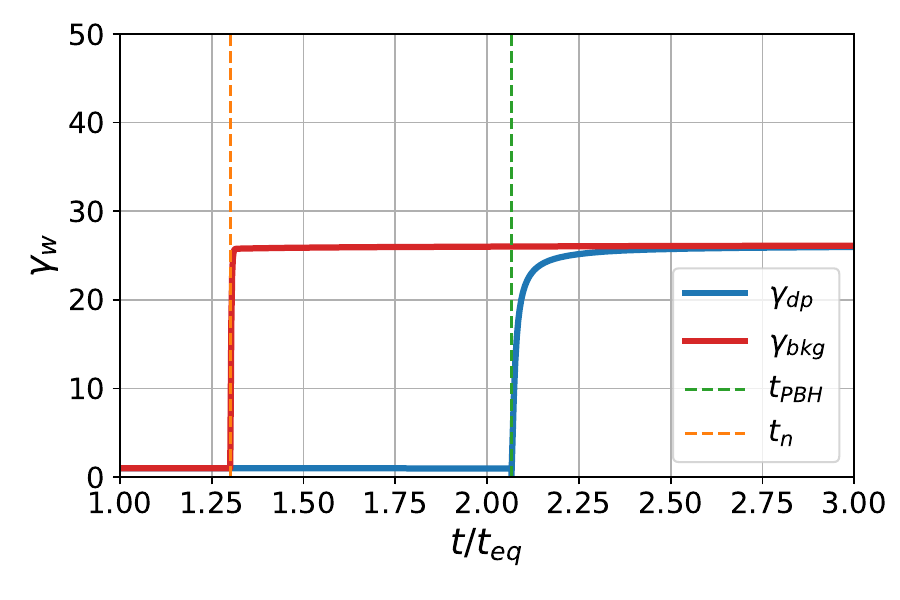}
\includegraphics[scale=0.5]{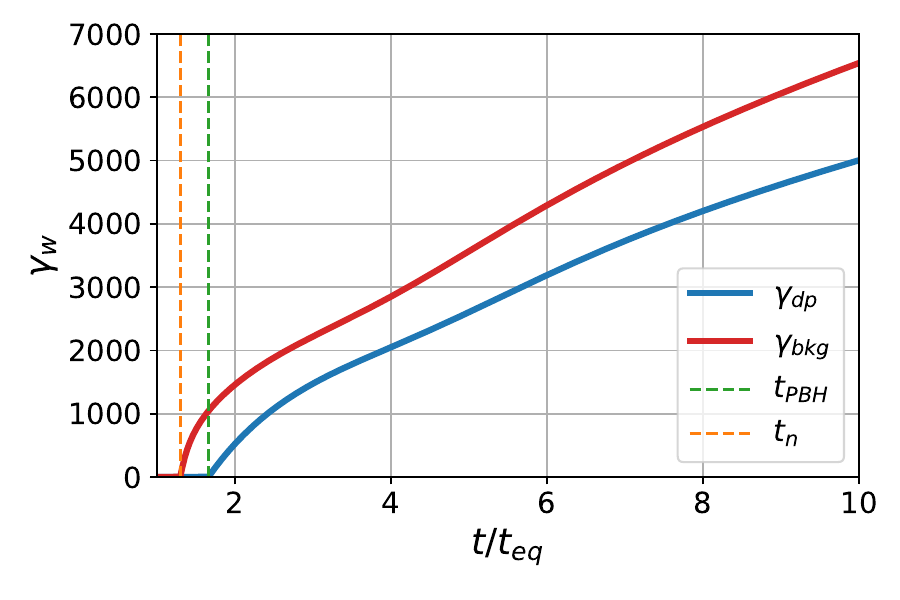}\\
\includegraphics[scale=0.5]{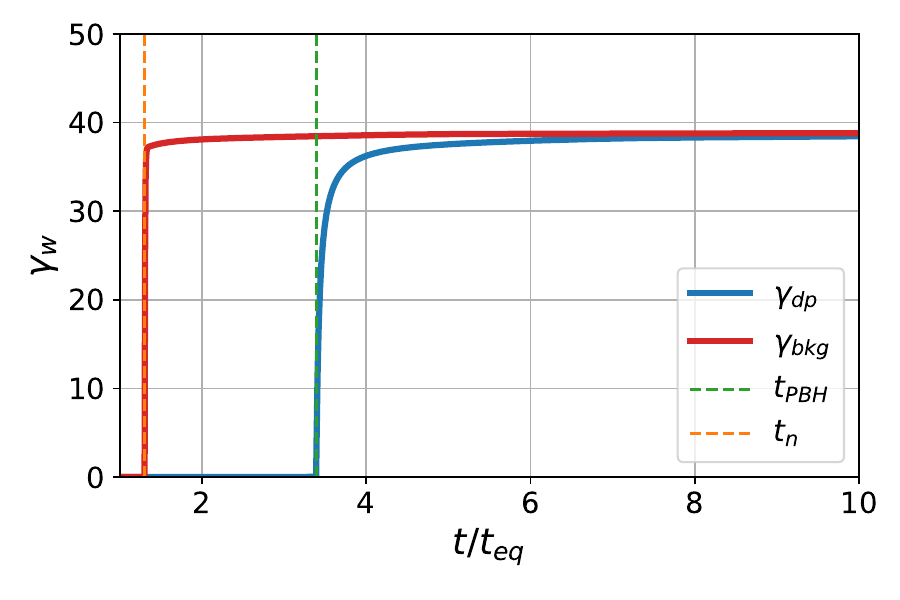}
\includegraphics[scale=0.5]{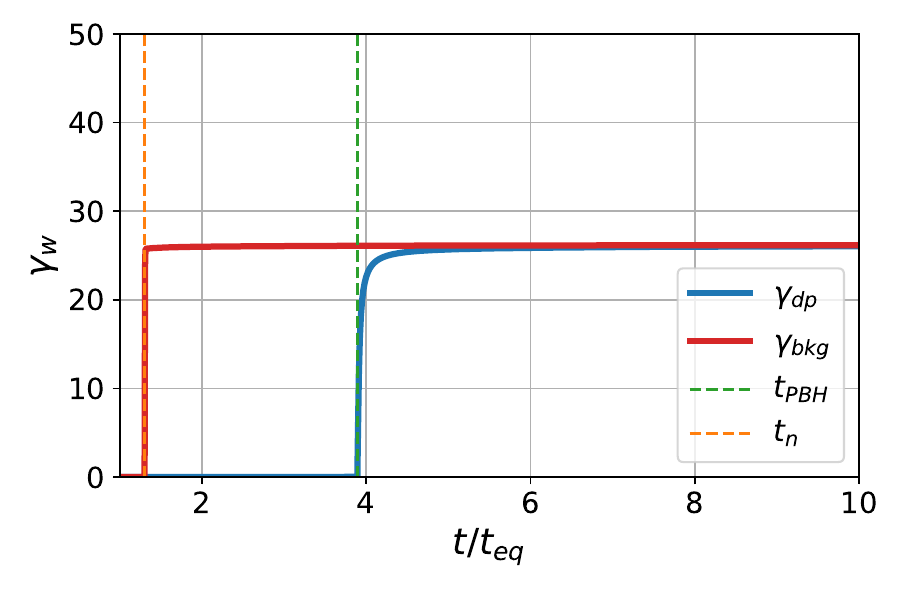}
\caption{The evolution of the lorentz factors ($\gamma_w$) associated with the bubble wall velocities for both the delayed patch and the background are shown here for the benchmark points mentioned in Table~\ref{tab:BP} in the following order: BP 1 (\textit{top left}), BP 2 (\textit{top right}), BP 3 (\textit{middle left}), BP 4 (\textit{middle right}), BP 5 (\textit{bottom left}), BP 6 (\textit{bottom right}).}
\label{fig:gamma}
\end{figure}

The friction terms in Eqs.~\eqref{eq:fric1} and \eqref{eq:fric2} depend explicitly on the lorentz factors ($\gamma_w$) associated with the bubble wall velocities. In the frictionless scenarios, the $\gamma_w$'s for both delayed patch and the background keep rising regardless of the nucleation rate profile. However, once the friction term is added, the $\gamma_w$'s reach a terminal velocity \cite{Ellis:2019oqb}. Simulations of a dynamical evolution of the pressure on the bubble wall in local thermal equilibrium and the effect on wall velocity can be found in Ref.~\cite{Laurent:2026jvx}. The evolution of $\gamma_w$'s in all of the six cases of Table.~\ref{tab:BP} are shown in Fig.~\ref{fig:gamma}.

\section{Estimation of $\gamma_c$ for `with-friction' benchmarks}
\label{sec:gamma_c}

Following Eq.~(105) of \cite{Gouttenoire:2021kjv}, we can express $\gamma_c$ as
\begin{equation}
    \gamma_c = \frac{T_n}{(\beta/H_n)\,H_n}\,.
\end{equation}

Assuming a vacuum energy dominated background, we approximate $H_n \simeq H(T_n) \sim \frac{T_n^2}{M_{\rm Pl}}$. This follows from the condition $\rho_v(T_{\rm eq}) = \rho_R(T_{\rm eq})$ with $T_{\rm eq} \simeq T_n$ at the cosmological scale. Substituting this we obtain
\begin{equation}
    \gamma_c = \frac{M_{\rm Pl}}{(\beta/H_n)\,T_n}\,.
\end{equation}

For the benchmark values $(\beta/H_n) = 20$ and $T_n = 10^2\,\text{GeV}$ (see Table~\ref{tab:BP}), we find $\gamma_c \sim 10^{15}$. In contrast, as shown in Fig.~\ref{fig:gamma} for all `with-friction' scenarios the terminal Lorentz factor satisfies $\gamma_t \sim 10^2 \ll \gamma_c$. This implies that the bubble walls reach their terminal velocity almost immediately after formation. Consequently, most of the vacuum energy is efficiently transferred to the plasma and sound waves and MHD turbulence dominate over the bubble collision contribution to the GW signal.

This also provides an approximate upper bound on the combination $(\beta/H_n)\,T_n$, below which plasma induced sources dominate over bubble collisions.

\bibliography{ref}

\begin{thebibliography}{84}%
\makeatletter
\providecommand \@ifxundefined [1]{%
 \@ifx{#1\undefined}
}%
\providecommand \@ifnum [1]{%
 \ifnum #1\expandafter \@firstoftwo
 \else \expandafter \@secondoftwo
 \fi
}%
\providecommand \@ifx [1]{%
 \ifx #1\expandafter \@firstoftwo
 \else \expandafter \@secondoftwo
 \fi
}%
\providecommand \natexlab [1]{#1}%
\providecommand \enquote  [1]{``#1''}%
\providecommand \bibnamefont  [1]{#1}%
\providecommand \bibfnamefont [1]{#1}%
\providecommand \citenamefont [1]{#1}%
\providecommand \href@noop [0]{\@secondoftwo}%
\providecommand \href [0]{\begingroup \@sanitize@url \@href}%
\providecommand \@href[1]{\@@startlink{#1}\@@href}%
\providecommand \@@href[1]{\endgroup#1\@@endlink}%
\providecommand \@sanitize@url [0]{\catcode `\\12\catcode `\$12\catcode
  `\&12\catcode `\#12\catcode `\^12\catcode `\_12\catcode `\%12\relax}%
\providecommand \@@startlink[1]{}%
\providecommand \@@endlink[0]{}%
\providecommand \url  [0]{\begingroup\@sanitize@url \@url }%
\providecommand \@url [1]{\endgroup\@href {#1}{\urlprefix }}%
\providecommand \urlprefix  [0]{URL }%
\providecommand \Eprint [0]{\href }%
\providecommand \doibase [0]{https://doi.org/}%
\providecommand \selectlanguage [0]{\@gobble}%
\providecommand \bibinfo  [0]{\@secondoftwo}%
\providecommand \bibfield  [0]{\@secondoftwo}%
\providecommand \translation [1]{[#1]}%
\providecommand \BibitemOpen [0]{}%
\providecommand \bibitemStop [0]{}%
\providecommand \bibitemNoStop [0]{.\EOS\space}%
\providecommand \EOS [0]{\spacefactor3000\relax}%
\providecommand \BibitemShut  [1]{\csname bibitem#1\endcsname}%
\let\auto@bib@innerbib\@empty
\bibitem [{\citenamefont {Hinshaw}\ \emph {et~al.}(2013)\citenamefont {Hinshaw}
  \emph {et~al.}}]{WMAP:2012nax}%
  \BibitemOpen
  \bibfield  {author} {\bibinfo {author} {\bibfnamefont {G.}~\bibnamefont
  {Hinshaw}} \emph {et~al.} (\bibinfo {collaboration} {WMAP}),\ }\bibfield
  {title} {\bibinfo {title} {{Nine-Year Wilkinson Microwave Anisotropy Probe
  (WMAP) Observations: Cosmological Parameter Results}},\ }\href
  {https://doi.org/10.1088/0067-0049/208/2/19} {\bibfield  {journal} {\bibinfo
  {journal} {Astrophys. J. Suppl.}\ }\textbf {\bibinfo {volume} {208}},\
  \bibinfo {pages} {19} (\bibinfo {year} {2013})},\ \Eprint
  {https://arxiv.org/abs/1212.5226} {arXiv:1212.5226 [astro-ph.CO]}
  \BibitemShut {NoStop}%
\bibitem [{\citenamefont {Aghanim}\ \emph {et~al.}(2020)\citenamefont {Aghanim}
  \emph {et~al.}}]{Planck:2018vyg}%
  \BibitemOpen
  \bibfield  {author} {\bibinfo {author} {\bibfnamefont {N.}~\bibnamefont
  {Aghanim}} \emph {et~al.} (\bibinfo {collaboration} {Planck}),\ }\bibfield
  {title} {\bibinfo {title} {{Planck 2018 results. VI. Cosmological
  parameters}},\ }\href {https://doi.org/10.1051/0004-6361/201833910}
  {\bibfield  {journal} {\bibinfo  {journal} {Astron. Astrophys.}\ }\textbf
  {\bibinfo {volume} {641}},\ \bibinfo {pages} {A6} (\bibinfo {year} {2020})},\
  \bibinfo {note} {[Erratum: Astron.Astrophys. 652, C4 (2021)]},\ \Eprint
  {https://arxiv.org/abs/1807.06209} {arXiv:1807.06209 [astro-ph.CO]}
  \BibitemShut {NoStop}%
\bibitem [{\citenamefont {Cyburt}\ \emph {et~al.}(2016)\citenamefont {Cyburt},
  \citenamefont {Fields}, \citenamefont {Olive},\ and\ \citenamefont
  {Yeh}}]{Cyburt:2015mya}%
  \BibitemOpen
  \bibfield  {author} {\bibinfo {author} {\bibfnamefont {R.~H.}\ \bibnamefont
  {Cyburt}}, \bibinfo {author} {\bibfnamefont {B.~D.}\ \bibnamefont {Fields}},
  \bibinfo {author} {\bibfnamefont {K.~A.}\ \bibnamefont {Olive}},\ and\
  \bibinfo {author} {\bibfnamefont {T.-H.}\ \bibnamefont {Yeh}},\ }\bibfield
  {title} {\bibinfo {title} {{Big Bang Nucleosynthesis: 2015}},\ }\href
  {https://doi.org/10.1103/RevModPhys.88.015004} {\bibfield  {journal}
  {\bibinfo  {journal} {Rev. Mod. Phys.}\ }\textbf {\bibinfo {volume} {88}},\
  \bibinfo {pages} {015004} (\bibinfo {year} {2016})},\ \Eprint
  {https://arxiv.org/abs/1505.01076} {arXiv:1505.01076 [astro-ph.CO]}
  \BibitemShut {NoStop}%
\bibitem [{\citenamefont {Pitrou}\ \emph {et~al.}(2018)\citenamefont {Pitrou},
  \citenamefont {Coc}, \citenamefont {Uzan},\ and\ \citenamefont
  {Vangioni}}]{Pitrou:2018cgg}%
  \BibitemOpen
  \bibfield  {author} {\bibinfo {author} {\bibfnamefont {C.}~\bibnamefont
  {Pitrou}}, \bibinfo {author} {\bibfnamefont {A.}~\bibnamefont {Coc}},
  \bibinfo {author} {\bibfnamefont {J.-P.}\ \bibnamefont {Uzan}},\ and\
  \bibinfo {author} {\bibfnamefont {E.}~\bibnamefont {Vangioni}},\ }\bibfield
  {title} {\bibinfo {title} {{Precision big bang nucleosynthesis with improved
  Helium-4 predictions}},\ }\href
  {https://doi.org/10.1016/j.physrep.2018.04.005} {\bibfield  {journal}
  {\bibinfo  {journal} {Phys. Rept.}\ }\textbf {\bibinfo {volume} {754}},\
  \bibinfo {pages} {1} (\bibinfo {year} {2018})},\ \Eprint
  {https://arxiv.org/abs/1801.08023} {arXiv:1801.08023 [astro-ph.CO]}
  \BibitemShut {NoStop}%
\bibitem [{\citenamefont {Abbott}\ \emph {et~al.}(2016)\citenamefont {Abbott}
  \emph {et~al.}}]{LIGOScientific:2016aoc}%
  \BibitemOpen
  \bibfield  {author} {\bibinfo {author} {\bibfnamefont {B.~P.}\ \bibnamefont
  {Abbott}} \emph {et~al.} (\bibinfo {collaboration} {LIGO Scientific,
  Virgo}),\ }\bibfield  {title} {\bibinfo {title} {{Observation of
  Gravitational Waves from a Binary Black Hole Merger}},\ }\href
  {https://doi.org/10.1103/PhysRevLett.116.061102} {\bibfield  {journal}
  {\bibinfo  {journal} {Phys. Rev. Lett.}\ }\textbf {\bibinfo {volume} {116}},\
  \bibinfo {pages} {061102} (\bibinfo {year} {2016})},\ \Eprint
  {https://arxiv.org/abs/1602.03837} {arXiv:1602.03837 [gr-qc]} \BibitemShut
  {NoStop}%
\bibitem [{\citenamefont {Abbott}\ \emph
  {et~al.}(2017{\natexlab{a}})\citenamefont {Abbott} \emph
  {et~al.}}]{LIGOScientific:2017vwq}%
  \BibitemOpen
  \bibfield  {author} {\bibinfo {author} {\bibfnamefont {B.~P.}\ \bibnamefont
  {Abbott}} \emph {et~al.} (\bibinfo {collaboration} {LIGO Scientific,
  Virgo}),\ }\bibfield  {title} {\bibinfo {title} {{GW170817: Observation of
  Gravitational Waves from a Binary Neutron Star Inspiral}},\ }\href
  {https://doi.org/10.1103/PhysRevLett.119.161101} {\bibfield  {journal}
  {\bibinfo  {journal} {Phys. Rev. Lett.}\ }\textbf {\bibinfo {volume} {119}},\
  \bibinfo {pages} {161101} (\bibinfo {year} {2017}{\natexlab{a}})},\ \Eprint
  {https://arxiv.org/abs/1710.05832} {arXiv:1710.05832 [gr-qc]} \BibitemShut
  {NoStop}%
\bibitem [{\citenamefont {Agazie}\ \emph
  {et~al.}(2023{\natexlab{a}})\citenamefont {Agazie} \emph
  {et~al.}}]{NANOGrav:2023gor}%
  \BibitemOpen
  \bibfield  {author} {\bibinfo {author} {\bibfnamefont {G.}~\bibnamefont
  {Agazie}} \emph {et~al.} (\bibinfo {collaboration} {NANOGrav}),\ }\bibfield
  {title} {\bibinfo {title} {{The NANOGrav 15 yr Data Set: Evidence for a
  Gravitational-wave Background}},\ }\href
  {https://doi.org/10.3847/2041-8213/acdac6} {\bibfield  {journal} {\bibinfo
  {journal} {Astrophys. J. Lett.}\ }\textbf {\bibinfo {volume} {951}},\
  \bibinfo {pages} {L8} (\bibinfo {year} {2023}{\natexlab{a}})},\ \Eprint
  {https://arxiv.org/abs/2306.16213} {arXiv:2306.16213 [astro-ph.HE]}
  \BibitemShut {NoStop}%
\bibitem [{\citenamefont {Agazie}\ \emph
  {et~al.}(2023{\natexlab{b}})\citenamefont {Agazie} \emph
  {et~al.}}]{NANOGrav:2023hde}%
  \BibitemOpen
  \bibfield  {author} {\bibinfo {author} {\bibfnamefont {G.}~\bibnamefont
  {Agazie}} \emph {et~al.} (\bibinfo {collaboration} {NANOGrav}),\ }\bibfield
  {title} {\bibinfo {title} {{The NANOGrav 15 yr Data Set: Observations and
  Timing of 68 Millisecond Pulsars}},\ }\href
  {https://doi.org/10.3847/2041-8213/acda9a} {\bibfield  {journal} {\bibinfo
  {journal} {Astrophys. J. Lett.}\ }\textbf {\bibinfo {volume} {951}},\
  \bibinfo {pages} {L9} (\bibinfo {year} {2023}{\natexlab{b}})},\ \Eprint
  {https://arxiv.org/abs/2306.16217} {arXiv:2306.16217 [astro-ph.HE]}
  \BibitemShut {NoStop}%
\bibitem [{\citenamefont {Antoniadis}\ \emph {et~al.}(2023)\citenamefont
  {Antoniadis} \emph {et~al.}}]{EPTA:2023fyk}%
  \BibitemOpen
  \bibfield  {author} {\bibinfo {author} {\bibfnamefont {J.}~\bibnamefont
  {Antoniadis}} \emph {et~al.} (\bibinfo {collaboration} {EPTA, InPTA:}),\
  }\bibfield  {title} {\bibinfo {title} {{The second data release from the
  European Pulsar Timing Array - III. Search for gravitational wave signals}},\
  }\href {https://doi.org/10.1051/0004-6361/202346844} {\bibfield  {journal}
  {\bibinfo  {journal} {Astron. Astrophys.}\ }\textbf {\bibinfo {volume}
  {678}},\ \bibinfo {pages} {A50} (\bibinfo {year} {2023})},\ \Eprint
  {https://arxiv.org/abs/2306.16214} {arXiv:2306.16214 [astro-ph.HE]}
  \BibitemShut {NoStop}%
\bibitem [{\citenamefont {Reardon}\ \emph {et~al.}(2023)\citenamefont {Reardon}
  \emph {et~al.}}]{Reardon:2023gzh}%
  \BibitemOpen
  \bibfield  {author} {\bibinfo {author} {\bibfnamefont {D.~J.}\ \bibnamefont
  {Reardon}} \emph {et~al.},\ }\bibfield  {title} {\bibinfo {title} {{Search
  for an Isotropic Gravitational-wave Background with the Parkes Pulsar Timing
  Array}},\ }\href {https://doi.org/10.3847/2041-8213/acdd02} {\bibfield
  {journal} {\bibinfo  {journal} {Astrophys. J. Lett.}\ }\textbf {\bibinfo
  {volume} {951}},\ \bibinfo {pages} {L6} (\bibinfo {year} {2023})},\ \Eprint
  {https://arxiv.org/abs/2306.16215} {arXiv:2306.16215 [astro-ph.HE]}
  \BibitemShut {NoStop}%
\bibitem [{\citenamefont {Xu}\ \emph {et~al.}(2023)\citenamefont {Xu} \emph
  {et~al.}}]{Xu:2023wog}%
  \BibitemOpen
  \bibfield  {author} {\bibinfo {author} {\bibfnamefont {H.}~\bibnamefont {Xu}}
  \emph {et~al.},\ }\bibfield  {title} {\bibinfo {title} {{Searching for the
  Nano-Hertz Stochastic Gravitational Wave Background with the Chinese Pulsar
  Timing Array Data Release I}},\ }\href
  {https://doi.org/10.1088/1674-4527/acdfa5} {\bibfield  {journal} {\bibinfo
  {journal} {Res. Astron. Astrophys.}\ }\textbf {\bibinfo {volume} {23}},\
  \bibinfo {pages} {075024} (\bibinfo {year} {2023})},\ \Eprint
  {https://arxiv.org/abs/2306.16216} {arXiv:2306.16216 [astro-ph.HE]}
  \BibitemShut {NoStop}%
\bibitem [{\citenamefont {Afzal}\ \emph {et~al.}(2023)\citenamefont {Afzal}
  \emph {et~al.}}]{NANOGrav:2023hvm}%
  \BibitemOpen
  \bibfield  {author} {\bibinfo {author} {\bibfnamefont {A.}~\bibnamefont
  {Afzal}} \emph {et~al.} (\bibinfo {collaboration} {NANOGrav}),\ }\bibfield
  {title} {\bibinfo {title} {{The NANOGrav 15 yr Data Set: Search for Signals
  from New Physics}},\ }\href {https://doi.org/10.3847/2041-8213/acdc91}
  {\bibfield  {journal} {\bibinfo  {journal} {Astrophys. J. Lett.}\ }\textbf
  {\bibinfo {volume} {951}},\ \bibinfo {pages} {L11} (\bibinfo {year}
  {2023})},\ \bibinfo {note} {[Erratum: Astrophys.J.Lett. 971, L27 (2024),
  Erratum: Astrophys.J. 971, L27 (2024)]},\ \Eprint
  {https://arxiv.org/abs/2306.16219} {arXiv:2306.16219 [astro-ph.HE]}
  \BibitemShut {NoStop}%
\bibitem [{\citenamefont {Saito}\ and\ \citenamefont
  {Yokoyama}(2009)}]{Saito:2008jc}%
  \BibitemOpen
  \bibfield  {author} {\bibinfo {author} {\bibfnamefont {R.}~\bibnamefont
  {Saito}}\ and\ \bibinfo {author} {\bibfnamefont {J.}~\bibnamefont
  {Yokoyama}},\ }\bibfield  {title} {\bibinfo {title} {{Gravitational wave
  background as a probe of the primordial black hole abundance}},\ }\href
  {https://doi.org/10.1103/PhysRevLett.102.161101} {\bibfield  {journal}
  {\bibinfo  {journal} {Phys. Rev. Lett.}\ }\textbf {\bibinfo {volume} {102}},\
  \bibinfo {pages} {161101} (\bibinfo {year} {2009})},\ \bibinfo {note}
  {[Erratum: Phys.Rev.Lett. 107, 069901 (2011)]},\ \Eprint
  {https://arxiv.org/abs/0812.4339} {arXiv:0812.4339 [astro-ph]} \BibitemShut
  {NoStop}%
\bibitem [{\citenamefont {Inomata}\ \emph {et~al.}(2020)\citenamefont
  {Inomata}, \citenamefont {Kawasaki}, \citenamefont {Mukaida}, \citenamefont
  {Terada},\ and\ \citenamefont {Yanagida}}]{Inomata:2020lmk}%
  \BibitemOpen
  \bibfield  {author} {\bibinfo {author} {\bibfnamefont {K.}~\bibnamefont
  {Inomata}}, \bibinfo {author} {\bibfnamefont {M.}~\bibnamefont {Kawasaki}},
  \bibinfo {author} {\bibfnamefont {K.}~\bibnamefont {Mukaida}}, \bibinfo
  {author} {\bibfnamefont {T.}~\bibnamefont {Terada}},\ and\ \bibinfo {author}
  {\bibfnamefont {T.~T.}\ \bibnamefont {Yanagida}},\ }\bibfield  {title}
  {\bibinfo {title} {{Gravitational Wave Production right after a Primordial
  Black Hole Evaporation}},\ }\href
  {https://doi.org/10.1103/PhysRevD.101.123533} {\bibfield  {journal} {\bibinfo
   {journal} {Phys. Rev. D}\ }\textbf {\bibinfo {volume} {101}},\ \bibinfo
  {pages} {123533} (\bibinfo {year} {2020})},\ \Eprint
  {https://arxiv.org/abs/2003.10455} {arXiv:2003.10455 [astro-ph.CO]}
  \BibitemShut {NoStop}%
\bibitem [{\citenamefont {Zel'dovich}\ and\ \citenamefont
  {Novikov}(1967)}]{Zeldovich:1967lct}%
  \BibitemOpen
  \bibfield  {author} {\bibinfo {author} {\bibfnamefont {Y.~B.}\ \bibnamefont
  {Zel'dovich}}\ and\ \bibinfo {author} {\bibfnamefont {I.~D.}\ \bibnamefont
  {Novikov}},\ }\bibfield  {title} {\bibinfo {title} {{The Hypothesis of Cores
  Retarded during Expansion and the Hot Cosmological Model}},\ }\href
  {https://doi.org/https://ui.adsabs.harvard.edu/abs/1966AZh....43..758Z/abstract}
  {\bibfield  {journal} {\bibinfo  {journal} {Sov. Astron.}\ }\textbf {\bibinfo
  {volume} {10}},\ \bibinfo {pages} {602} (\bibinfo {year} {1967})}\BibitemShut
  {NoStop}%
\bibitem [{\citenamefont {Hawking}(1971)}]{Hawking:1971ei}%
  \BibitemOpen
  \bibfield  {author} {\bibinfo {author} {\bibfnamefont {S.}~\bibnamefont
  {Hawking}},\ }\bibfield  {title} {\bibinfo {title} {{Gravitationally
  collapsed objects of very low mass}},\ }\href
  {https://doi.org/10.1093/mnras/152.1.75} {\bibfield  {journal} {\bibinfo
  {journal} {Mon. Not. Roy. Astron. Soc.}\ }\textbf {\bibinfo {volume} {152}},\
  \bibinfo {pages} {75} (\bibinfo {year} {1971})}\BibitemShut {NoStop}%
\bibitem [{\citenamefont {Carr}\ and\ \citenamefont
  {Hawking}(1974)}]{Carr:1974nx}%
  \BibitemOpen
  \bibfield  {author} {\bibinfo {author} {\bibfnamefont {B.~J.}\ \bibnamefont
  {Carr}}\ and\ \bibinfo {author} {\bibfnamefont {S.~W.}\ \bibnamefont
  {Hawking}},\ }\bibfield  {title} {\bibinfo {title} {{Black holes in the early
  Universe}},\ }\href {https://doi.org/10.1093/mnras/168.2.399} {\bibfield
  {journal} {\bibinfo  {journal} {Mon. Not. Roy. Astron. Soc.}\ }\textbf
  {\bibinfo {volume} {168}},\ \bibinfo {pages} {399} (\bibinfo {year}
  {1974})}\BibitemShut {NoStop}%
\bibitem [{\citenamefont {Carr}(1975)}]{Carr:1975qj}%
  \BibitemOpen
  \bibfield  {author} {\bibinfo {author} {\bibfnamefont {B.~J.}\ \bibnamefont
  {Carr}},\ }\bibfield  {title} {\bibinfo {title} {{The Primordial black hole
  mass spectrum}},\ }\href {https://doi.org/10.1086/153853} {\bibfield
  {journal} {\bibinfo  {journal} {Astrophys. J.}\ }\textbf {\bibinfo {volume}
  {201}},\ \bibinfo {pages} {1} (\bibinfo {year} {1975})}\BibitemShut {NoStop}%
\bibitem [{\citenamefont {Hawking}(1989)}]{Hawking:1987bn}%
  \BibitemOpen
  \bibfield  {author} {\bibinfo {author} {\bibfnamefont {S.~W.}\ \bibnamefont
  {Hawking}},\ }\bibfield  {title} {\bibinfo {title} {{Black Holes From Cosmic
  Strings}},\ }\href {https://doi.org/10.1016/0370-2693(89)90206-2} {\bibfield
  {journal} {\bibinfo  {journal} {Phys. Lett. B}\ }\textbf {\bibinfo {volume}
  {231}},\ \bibinfo {pages} {237} (\bibinfo {year} {1989})}\BibitemShut
  {NoStop}%
\bibitem [{\citenamefont {Polnarev}\ and\ \citenamefont
  {Zembowicz}(1991)}]{Polnarev:1988dh}%
  \BibitemOpen
  \bibfield  {author} {\bibinfo {author} {\bibfnamefont {A.}~\bibnamefont
  {Polnarev}}\ and\ \bibinfo {author} {\bibfnamefont {R.}~\bibnamefont
  {Zembowicz}},\ }\bibfield  {title} {\bibinfo {title} {{Formation of
  Primordial Black Holes by Cosmic Strings}},\ }\href
  {https://doi.org/10.1103/PhysRevD.43.1106} {\bibfield  {journal} {\bibinfo
  {journal} {Phys. Rev. D}\ }\textbf {\bibinfo {volume} {43}},\ \bibinfo
  {pages} {1106} (\bibinfo {year} {1991})}\BibitemShut {NoStop}%
\bibitem [{\citenamefont {Jenkins}\ and\ \citenamefont
  {Sakellariadou}(2020)}]{Jenkins:2020ctp}%
  \BibitemOpen
  \bibfield  {author} {\bibinfo {author} {\bibfnamefont {A.~C.}\ \bibnamefont
  {Jenkins}}\ and\ \bibinfo {author} {\bibfnamefont {M.}~\bibnamefont
  {Sakellariadou}},\ }\bibfield  {title} {\bibinfo {title} {{Primordial black
  holes from cusp collapse on cosmic strings}},\ }\href@noop {} {\  (\bibinfo
  {year} {2020})},\ \Eprint {https://arxiv.org/abs/2006.16249}
  {arXiv:2006.16249 [astro-ph.CO]} \BibitemShut {NoStop}%
\bibitem [{\citenamefont {Deng}\ \emph {et~al.}(2017)\citenamefont {Deng},
  \citenamefont {Garriga},\ and\ \citenamefont {Vilenkin}}]{Deng:2016vzb}%
  \BibitemOpen
  \bibfield  {author} {\bibinfo {author} {\bibfnamefont {H.}~\bibnamefont
  {Deng}}, \bibinfo {author} {\bibfnamefont {J.}~\bibnamefont {Garriga}},\ and\
  \bibinfo {author} {\bibfnamefont {A.}~\bibnamefont {Vilenkin}},\ }\bibfield
  {title} {\bibinfo {title} {{Primordial black hole and wormhole formation by
  domain walls}},\ }\href {https://doi.org/10.1088/1475-7516/2017/04/050}
  {\bibfield  {journal} {\bibinfo  {journal} {JCAP}\ }\textbf {\bibinfo
  {volume} {04}},\ \bibinfo {pages} {050}},\ \Eprint
  {https://arxiv.org/abs/1612.03753} {arXiv:1612.03753 [gr-qc]} \BibitemShut
  {NoStop}%
\bibitem [{\citenamefont {Garriga}\ \emph {et~al.}(2016)\citenamefont
  {Garriga}, \citenamefont {Vilenkin},\ and\ \citenamefont
  {Zhang}}]{Garriga:2015fdk}%
  \BibitemOpen
  \bibfield  {author} {\bibinfo {author} {\bibfnamefont {J.}~\bibnamefont
  {Garriga}}, \bibinfo {author} {\bibfnamefont {A.}~\bibnamefont {Vilenkin}},\
  and\ \bibinfo {author} {\bibfnamefont {J.}~\bibnamefont {Zhang}},\ }\bibfield
   {title} {\bibinfo {title} {{Black holes and the multiverse}},\ }\href
  {https://doi.org/10.1088/1475-7516/2016/02/064} {\bibfield  {journal}
  {\bibinfo  {journal} {JCAP}\ }\textbf {\bibinfo {volume} {02}},\ \bibinfo
  {pages} {064}},\ \Eprint {https://arxiv.org/abs/1512.01819} {arXiv:1512.01819
  [hep-th]} \BibitemShut {NoStop}%
\bibitem [{\citenamefont {Liu}\ \emph {et~al.}(2020)\citenamefont {Liu},
  \citenamefont {Guo},\ and\ \citenamefont {Cai}}]{Liu:2019lul}%
  \BibitemOpen
  \bibfield  {author} {\bibinfo {author} {\bibfnamefont {J.}~\bibnamefont
  {Liu}}, \bibinfo {author} {\bibfnamefont {Z.-K.}\ \bibnamefont {Guo}},\ and\
  \bibinfo {author} {\bibfnamefont {R.-G.}\ \bibnamefont {Cai}},\ }\bibfield
  {title} {\bibinfo {title} {{Primordial Black Holes from Cosmic Domain
  Walls}},\ }\href {https://doi.org/10.1103/PhysRevD.101.023513} {\bibfield
  {journal} {\bibinfo  {journal} {Phys. Rev. D}\ }\textbf {\bibinfo {volume}
  {101}},\ \bibinfo {pages} {023513} (\bibinfo {year} {2020})},\ \Eprint
  {https://arxiv.org/abs/1908.02662} {arXiv:1908.02662 [astro-ph.CO]}
  \BibitemShut {NoStop}%
\bibitem [{\citenamefont {Zeldovich}\ \emph {et~al.}(1974)\citenamefont
  {Zeldovich}, \citenamefont {Kobzarev},\ and\ \citenamefont
  {Okun}}]{Zeldovich:1974uw}%
  \BibitemOpen
  \bibfield  {author} {\bibinfo {author} {\bibfnamefont {Y.~B.}\ \bibnamefont
  {Zeldovich}}, \bibinfo {author} {\bibfnamefont {I.~Y.}\ \bibnamefont
  {Kobzarev}},\ and\ \bibinfo {author} {\bibfnamefont {L.~B.}\ \bibnamefont
  {Okun}},\ }\bibfield  {title} {\bibinfo {title} {{Cosmological Consequences
  of the Spontaneous Breakdown of Discrete Symmetry}},\ }\href
  {https://doi.org/http://jetp.ras.ru/cgi-bin/dn/e_040_01_0001.pdf} {\bibfield
  {journal} {\bibinfo  {journal} {Zh. Eksp. Teor. Fiz.}\ }\textbf {\bibinfo
  {volume} {67}},\ \bibinfo {pages} {3} (\bibinfo {year} {1974})}\BibitemShut
  {NoStop}%
\bibitem [{\citenamefont {Ge}(2020)}]{Ge:2019ihf}%
  \BibitemOpen
  \bibfield  {author} {\bibinfo {author} {\bibfnamefont {S.}~\bibnamefont
  {Ge}},\ }\bibfield  {title} {\bibinfo {title} {{Sublunar-Mass Primordial
  Black Holes from Closed Axion Domain Walls}},\ }\href
  {https://doi.org/10.1016/j.dark.2019.100440} {\bibfield  {journal} {\bibinfo
  {journal} {Phys. Dark Univ.}\ }\textbf {\bibinfo {volume} {27}},\ \bibinfo
  {pages} {100440} (\bibinfo {year} {2020})},\ \Eprint
  {https://arxiv.org/abs/1905.12182} {arXiv:1905.12182 [hep-ph]} \BibitemShut
  {NoStop}%
\bibitem [{\citenamefont {Crawford}\ and\ \citenamefont
  {Schramm}(1982)}]{Crawford:1982yz}%
  \BibitemOpen
  \bibfield  {author} {\bibinfo {author} {\bibfnamefont {M.}~\bibnamefont
  {Crawford}}\ and\ \bibinfo {author} {\bibfnamefont {D.~N.}\ \bibnamefont
  {Schramm}},\ }\bibfield  {title} {\bibinfo {title} {{Spontaneous Generation
  of Density Perturbations in the Early Universe}},\ }\href
  {https://doi.org/10.1038/298538a0} {\bibfield  {journal} {\bibinfo  {journal}
  {Nature}\ }\textbf {\bibinfo {volume} {298}},\ \bibinfo {pages} {538}
  (\bibinfo {year} {1982})}\BibitemShut {NoStop}%
\bibitem [{\citenamefont {Kodama}\ \emph {et~al.}(1982)\citenamefont {Kodama},
  \citenamefont {Sasaki},\ and\ \citenamefont {Sato}}]{Kodama:1982sf}%
  \BibitemOpen
  \bibfield  {author} {\bibinfo {author} {\bibfnamefont {H.}~\bibnamefont
  {Kodama}}, \bibinfo {author} {\bibfnamefont {M.}~\bibnamefont {Sasaki}},\
  and\ \bibinfo {author} {\bibfnamefont {K.}~\bibnamefont {Sato}},\ }\bibfield
  {title} {\bibinfo {title} {{Abundance of Primordial Holes Produced by
  Cosmological First Order Phase Transition}},\ }\href
  {https://doi.org/10.1143/PTP.68.1979} {\bibfield  {journal} {\bibinfo
  {journal} {Prog. Theor. Phys.}\ }\textbf {\bibinfo {volume} {68}},\ \bibinfo
  {pages} {1979} (\bibinfo {year} {1982})}\BibitemShut {NoStop}%
\bibitem [{\citenamefont {Hawking}\ \emph {et~al.}(1982)\citenamefont
  {Hawking}, \citenamefont {Moss},\ and\ \citenamefont
  {Stewart}}]{Hawking:1982ga}%
  \BibitemOpen
  \bibfield  {author} {\bibinfo {author} {\bibfnamefont {S.~W.}\ \bibnamefont
  {Hawking}}, \bibinfo {author} {\bibfnamefont {I.~G.}\ \bibnamefont {Moss}},\
  and\ \bibinfo {author} {\bibfnamefont {J.~M.}\ \bibnamefont {Stewart}},\
  }\bibfield  {title} {\bibinfo {title} {{Bubble Collisions in the Very Early
  Universe}},\ }\href {https://doi.org/10.1103/PhysRevD.26.2681} {\bibfield
  {journal} {\bibinfo  {journal} {Phys. Rev. D}\ }\textbf {\bibinfo {volume}
  {26}},\ \bibinfo {pages} {2681} (\bibinfo {year} {1982})}\BibitemShut
  {NoStop}%
\bibitem [{\citenamefont {Moss}(1994)}]{Moss:1994pi}%
  \BibitemOpen
  \bibfield  {author} {\bibinfo {author} {\bibfnamefont {I.~G.}\ \bibnamefont
  {Moss}},\ }\bibfield  {title} {\bibinfo {title} {{Black hole formation from
  colliding bubbles}},\ }\href@noop {} {\  (\bibinfo {year} {1994})},\ \Eprint
  {https://arxiv.org/abs/gr-qc/9405045} {arXiv:gr-qc/9405045} \BibitemShut
  {NoStop}%
\bibitem [{\citenamefont {Freivogel}\ \emph {et~al.}(2007)\citenamefont
  {Freivogel}, \citenamefont {Horowitz},\ and\ \citenamefont
  {Shenker}}]{Freivogel:2007fx}%
  \BibitemOpen
  \bibfield  {author} {\bibinfo {author} {\bibfnamefont {B.}~\bibnamefont
  {Freivogel}}, \bibinfo {author} {\bibfnamefont {G.~T.}\ \bibnamefont
  {Horowitz}},\ and\ \bibinfo {author} {\bibfnamefont {S.}~\bibnamefont
  {Shenker}},\ }\bibfield  {title} {\bibinfo {title} {{Colliding with a
  crunching bubble}},\ }\href {https://doi.org/10.1088/1126-6708/2007/05/090}
  {\bibfield  {journal} {\bibinfo  {journal} {JHEP}\ }\textbf {\bibinfo
  {volume} {05}},\ \bibinfo {pages} {090}},\ \Eprint
  {https://arxiv.org/abs/hep-th/0703146} {arXiv:hep-th/0703146} \BibitemShut
  {NoStop}%
\bibitem [{\citenamefont {Jedamzik}\ and\ \citenamefont
  {Niemeyer}(1999)}]{Jedamzik:1999am}%
  \BibitemOpen
  \bibfield  {author} {\bibinfo {author} {\bibfnamefont {K.}~\bibnamefont
  {Jedamzik}}\ and\ \bibinfo {author} {\bibfnamefont {J.~C.}\ \bibnamefont
  {Niemeyer}},\ }\bibfield  {title} {\bibinfo {title} {{Primordial black hole
  formation during first order phase transitions}},\ }\href
  {https://doi.org/10.1103/PhysRevD.59.124014} {\bibfield  {journal} {\bibinfo
  {journal} {Phys. Rev. D}\ }\textbf {\bibinfo {volume} {59}},\ \bibinfo
  {pages} {124014} (\bibinfo {year} {1999})},\ \Eprint
  {https://arxiv.org/abs/astro-ph/9901293} {arXiv:astro-ph/9901293}
  \BibitemShut {NoStop}%
\bibitem [{\citenamefont {Liu}\ \emph {et~al.}(2022)\citenamefont {Liu},
  \citenamefont {Bian}, \citenamefont {Cai}, \citenamefont {Guo},\ and\
  \citenamefont {Wang}}]{Liu:2021svg}%
  \BibitemOpen
  \bibfield  {author} {\bibinfo {author} {\bibfnamefont {J.}~\bibnamefont
  {Liu}}, \bibinfo {author} {\bibfnamefont {L.}~\bibnamefont {Bian}}, \bibinfo
  {author} {\bibfnamefont {R.-G.}\ \bibnamefont {Cai}}, \bibinfo {author}
  {\bibfnamefont {Z.-K.}\ \bibnamefont {Guo}},\ and\ \bibinfo {author}
  {\bibfnamefont {S.-J.}\ \bibnamefont {Wang}},\ }\bibfield  {title} {\bibinfo
  {title} {{Primordial black hole production during first-order phase
  transitions}},\ }\href {https://doi.org/10.1103/PhysRevD.105.L021303}
  {\bibfield  {journal} {\bibinfo  {journal} {Phys. Rev. D}\ }\textbf {\bibinfo
  {volume} {105}},\ \bibinfo {pages} {L021303} (\bibinfo {year} {2022})},\
  \Eprint {https://arxiv.org/abs/2106.05637} {arXiv:2106.05637 [astro-ph.CO]}
  \BibitemShut {NoStop}%
\bibitem [{\citenamefont {Gouttenoire}\ and\ \citenamefont
  {Volansky}(2024)}]{Gouttenoire:2023naa}%
  \BibitemOpen
  \bibfield  {author} {\bibinfo {author} {\bibfnamefont {Y.}~\bibnamefont
  {Gouttenoire}}\ and\ \bibinfo {author} {\bibfnamefont {T.}~\bibnamefont
  {Volansky}},\ }\bibfield  {title} {\bibinfo {title} {{Primordial black holes
  from supercooled phase transitions}},\ }\href
  {https://doi.org/10.1103/PhysRevD.110.043514} {\bibfield  {journal} {\bibinfo
   {journal} {Phys. Rev. D}\ }\textbf {\bibinfo {volume} {110}},\ \bibinfo
  {pages} {043514} (\bibinfo {year} {2024})},\ \Eprint
  {https://arxiv.org/abs/2305.04942} {arXiv:2305.04942 [hep-ph]} \BibitemShut
  {NoStop}%
\bibitem [{\citenamefont {Lewicki}\ \emph {et~al.}(2023)\citenamefont
  {Lewicki}, \citenamefont {Toczek},\ and\ \citenamefont
  {Vaskonen}}]{Lewicki:2023ioy}%
  \BibitemOpen
  \bibfield  {author} {\bibinfo {author} {\bibfnamefont {M.}~\bibnamefont
  {Lewicki}}, \bibinfo {author} {\bibfnamefont {P.}~\bibnamefont {Toczek}},\
  and\ \bibinfo {author} {\bibfnamefont {V.}~\bibnamefont {Vaskonen}},\
  }\bibfield  {title} {\bibinfo {title} {{Primordial black holes from strong
  first-order phase transitions}},\ }\href
  {https://doi.org/10.1007/JHEP09(2023)092} {\bibfield  {journal} {\bibinfo
  {journal} {JHEP}\ }\textbf {\bibinfo {volume} {09}},\ \bibinfo {pages}
  {092}},\ \Eprint {https://arxiv.org/abs/2305.04924} {arXiv:2305.04924
  [astro-ph.CO]} \BibitemShut {NoStop}%
\bibitem [{\citenamefont {Kanemura}\ \emph {et~al.}(2024)\citenamefont
  {Kanemura}, \citenamefont {Tanaka},\ and\ \citenamefont
  {Xie}}]{Kanemura:2024pae}%
  \BibitemOpen
  \bibfield  {author} {\bibinfo {author} {\bibfnamefont {S.}~\bibnamefont
  {Kanemura}}, \bibinfo {author} {\bibfnamefont {M.}~\bibnamefont {Tanaka}},\
  and\ \bibinfo {author} {\bibfnamefont {K.-P.}\ \bibnamefont {Xie}},\
  }\bibfield  {title} {\bibinfo {title} {{Primordial black holes from slow
  phase transitions: a model-building perspective}},\ }\href
  {https://doi.org/10.1007/JHEP06(2024)036} {\bibfield  {journal} {\bibinfo
  {journal} {JHEP}\ }\textbf {\bibinfo {volume} {06}},\ \bibinfo {pages}
  {036}},\ \Eprint {https://arxiv.org/abs/2404.00646} {arXiv:2404.00646
  [hep-ph]} \BibitemShut {NoStop}%
\bibitem [{\citenamefont {Cotner}\ and\ \citenamefont
  {Kusenko}(2017)}]{Cotner:2016cvr}%
  \BibitemOpen
  \bibfield  {author} {\bibinfo {author} {\bibfnamefont {E.}~\bibnamefont
  {Cotner}}\ and\ \bibinfo {author} {\bibfnamefont {A.}~\bibnamefont
  {Kusenko}},\ }\bibfield  {title} {\bibinfo {title} {{Primordial black holes
  from supersymmetry in the early universe}},\ }\href
  {https://doi.org/10.1103/PhysRevLett.119.031103} {\bibfield  {journal}
  {\bibinfo  {journal} {Phys. Rev. Lett.}\ }\textbf {\bibinfo {volume} {119}},\
  \bibinfo {pages} {031103} (\bibinfo {year} {2017})},\ \Eprint
  {https://arxiv.org/abs/1612.02529} {arXiv:1612.02529 [astro-ph.CO]}
  \BibitemShut {NoStop}%
\bibitem [{\citenamefont {Kawana}\ and\ \citenamefont
  {Xie}(2022)}]{Kawana:2021tde}%
  \BibitemOpen
  \bibfield  {author} {\bibinfo {author} {\bibfnamefont {K.}~\bibnamefont
  {Kawana}}\ and\ \bibinfo {author} {\bibfnamefont {K.-P.}\ \bibnamefont
  {Xie}},\ }\bibfield  {title} {\bibinfo {title} {{Primordial black holes from
  a cosmic phase transition: The collapse of Fermi-balls}},\ }\href
  {https://doi.org/10.1016/j.physletb.2021.136791} {\bibfield  {journal}
  {\bibinfo  {journal} {Phys. Lett. B}\ }\textbf {\bibinfo {volume} {824}},\
  \bibinfo {pages} {136791} (\bibinfo {year} {2022})},\ \Eprint
  {https://arxiv.org/abs/2106.00111} {arXiv:2106.00111 [astro-ph.CO]}
  \BibitemShut {NoStop}%
\bibitem [{\citenamefont {Baker}\ \emph
  {et~al.}(2025{\natexlab{a}})\citenamefont {Baker}, \citenamefont {Breitbach},
  \citenamefont {Kopp},\ and\ \citenamefont {Mittnacht}}]{Baker:2021nyl}%
  \BibitemOpen
  \bibfield  {author} {\bibinfo {author} {\bibfnamefont {M.~J.}\ \bibnamefont
  {Baker}}, \bibinfo {author} {\bibfnamefont {M.}~\bibnamefont {Breitbach}},
  \bibinfo {author} {\bibfnamefont {J.}~\bibnamefont {Kopp}},\ and\ \bibinfo
  {author} {\bibfnamefont {L.}~\bibnamefont {Mittnacht}},\ }\bibfield  {title}
  {\bibinfo {title} {{Primordial black holes from first-order cosmological
  phase transitions}},\ }\href {https://doi.org/10.1016/j.physletb.2025.139625}
  {\bibfield  {journal} {\bibinfo  {journal} {Phys. Lett. B}\ }\textbf
  {\bibinfo {volume} {868}},\ \bibinfo {pages} {139625} (\bibinfo {year}
  {2025}{\natexlab{a}})},\ \Eprint {https://arxiv.org/abs/2105.07481}
  {arXiv:2105.07481 [astro-ph.CO]} \BibitemShut {NoStop}%
\bibitem [{\citenamefont {Baker}\ \emph
  {et~al.}(2025{\natexlab{b}})\citenamefont {Baker}, \citenamefont {Breitbach},
  \citenamefont {Kopp},\ and\ \citenamefont {Mittnacht}}]{Baker:2021sno}%
  \BibitemOpen
  \bibfield  {author} {\bibinfo {author} {\bibfnamefont {M.~J.}\ \bibnamefont
  {Baker}}, \bibinfo {author} {\bibfnamefont {M.}~\bibnamefont {Breitbach}},
  \bibinfo {author} {\bibfnamefont {J.}~\bibnamefont {Kopp}},\ and\ \bibinfo
  {author} {\bibfnamefont {L.}~\bibnamefont {Mittnacht}},\ }\bibfield  {title}
  {\bibinfo {title} {{Detailed calculation of primordial black hole formation
  during first-order cosmological phase transitions}},\ }\href
  {https://doi.org/10.1103/PhysRevD.111.063544} {\bibfield  {journal} {\bibinfo
   {journal} {Phys. Rev. D}\ }\textbf {\bibinfo {volume} {111}},\ \bibinfo
  {pages} {063544} (\bibinfo {year} {2025}{\natexlab{b}})},\ \Eprint
  {https://arxiv.org/abs/2110.00005} {arXiv:2110.00005 [astro-ph.CO]}
  \BibitemShut {NoStop}%
\bibitem [{\citenamefont {Hashino}\ \emph {et~al.}(2022)\citenamefont
  {Hashino}, \citenamefont {Kanemura},\ and\ \citenamefont
  {Takahashi}}]{Hashino:2021qoq}%
  \BibitemOpen
  \bibfield  {author} {\bibinfo {author} {\bibfnamefont {K.}~\bibnamefont
  {Hashino}}, \bibinfo {author} {\bibfnamefont {S.}~\bibnamefont {Kanemura}},\
  and\ \bibinfo {author} {\bibfnamefont {T.}~\bibnamefont {Takahashi}},\
  }\bibfield  {title} {\bibinfo {title} {{Primordial black holes as a probe of
  strongly first-order electroweak phase transition}},\ }\href
  {https://doi.org/10.1016/j.physletb.2022.137261} {\bibfield  {journal}
  {\bibinfo  {journal} {Phys. Lett. B}\ }\textbf {\bibinfo {volume} {833}},\
  \bibinfo {pages} {137261} (\bibinfo {year} {2022})},\ \Eprint
  {https://arxiv.org/abs/2111.13099} {arXiv:2111.13099 [hep-ph]} \BibitemShut
  {NoStop}%
\bibitem [{\citenamefont {Banerjee}\ \emph {et~al.}(2025)\citenamefont
  {Banerjee}, \citenamefont {Rescigno},\ and\ \citenamefont
  {Salvio}}]{Banerjee:2024cwv}%
  \BibitemOpen
  \bibfield  {author} {\bibinfo {author} {\bibfnamefont {I.~K.}\ \bibnamefont
  {Banerjee}}, \bibinfo {author} {\bibfnamefont {F.}~\bibnamefont {Rescigno}},\
  and\ \bibinfo {author} {\bibfnamefont {A.}~\bibnamefont {Salvio}},\
  }\bibfield  {title} {\bibinfo {title} {{Primordial black holes (as dark
  matter) from the supercooled phase transitions with radiative symmetry
  breaking}},\ }\href {https://doi.org/10.1088/1475-7516/2025/07/007}
  {\bibfield  {journal} {\bibinfo  {journal} {JCAP}\ }\textbf {\bibinfo
  {volume} {07}},\ \bibinfo {pages} {007}},\ \Eprint
  {https://arxiv.org/abs/2412.06889} {arXiv:2412.06889 [hep-ph]} \BibitemShut
  {NoStop}%
\bibitem [{\citenamefont {Musco}(2019)}]{Musco:2018rwt}%
  \BibitemOpen
  \bibfield  {author} {\bibinfo {author} {\bibfnamefont {I.}~\bibnamefont
  {Musco}},\ }\bibfield  {title} {\bibinfo {title} {{Threshold for primordial
  black holes: Dependence on the shape of the cosmological perturbations}},\
  }\href {https://doi.org/10.1103/PhysRevD.100.123524} {\bibfield  {journal}
  {\bibinfo  {journal} {Phys. Rev. D}\ }\textbf {\bibinfo {volume} {100}},\
  \bibinfo {pages} {123524} (\bibinfo {year} {2019})},\ \Eprint
  {https://arxiv.org/abs/1809.02127} {arXiv:1809.02127 [gr-qc]} \BibitemShut
  {NoStop}%
\bibitem [{\citenamefont {Escriv{\`a}}\ \emph {et~al.}(2020)\citenamefont
  {Escriv{\`a}}, \citenamefont {Germani},\ and\ \citenamefont
  {Sheth}}]{Escriva:2019phb}%
  \BibitemOpen
  \bibfield  {author} {\bibinfo {author} {\bibfnamefont {A.}~\bibnamefont
  {Escriv{\`a}}}, \bibinfo {author} {\bibfnamefont {C.}~\bibnamefont
  {Germani}},\ and\ \bibinfo {author} {\bibfnamefont {R.~K.}\ \bibnamefont
  {Sheth}},\ }\bibfield  {title} {\bibinfo {title} {{Universal threshold for
  primordial black hole formation}},\ }\href
  {https://doi.org/10.1103/PhysRevD.101.044022} {\bibfield  {journal} {\bibinfo
   {journal} {Phys. Rev. D}\ }\textbf {\bibinfo {volume} {101}},\ \bibinfo
  {pages} {044022} (\bibinfo {year} {2020})},\ \Eprint
  {https://arxiv.org/abs/1907.13311} {arXiv:1907.13311 [gr-qc]} \BibitemShut
  {NoStop}%
\bibitem [{\citenamefont {Ning}\ \emph {et~al.}(2026)\citenamefont {Ning},
  \citenamefont {Zeng}, \citenamefont {Cai},\ and\ \citenamefont
  {Wang}}]{Ning:2026nfs}%
  \BibitemOpen
  \bibfield  {author} {\bibinfo {author} {\bibfnamefont {Z.}~\bibnamefont
  {Ning}}, \bibinfo {author} {\bibfnamefont {X.-X.}\ \bibnamefont {Zeng}},
  \bibinfo {author} {\bibfnamefont {R.-G.}\ \bibnamefont {Cai}},\ and\ \bibinfo
  {author} {\bibfnamefont {S.-J.}\ \bibnamefont {Wang}},\ }\bibfield  {title}
  {\bibinfo {title} {{Numerical simulations of primordial black hole formation
  via delayed first-order phase transitions}},\ }\href@noop {} {\  (\bibinfo
  {year} {2026})},\ \Eprint {https://arxiv.org/abs/2601.21878}
  {arXiv:2601.21878 [gr-qc]} \BibitemShut {NoStop}%
\bibitem [{\citenamefont {Franciolini}\ \emph {et~al.}(2026)\citenamefont
  {Franciolini}, \citenamefont {Gouttenoire},\ and\ \citenamefont
  {Jinno}}]{Franciolini:2025ztf}%
  \BibitemOpen
  \bibfield  {author} {\bibinfo {author} {\bibfnamefont {G.}~\bibnamefont
  {Franciolini}}, \bibinfo {author} {\bibfnamefont {Y.}~\bibnamefont
  {Gouttenoire}},\ and\ \bibinfo {author} {\bibfnamefont {R.}~\bibnamefont
  {Jinno}},\ }\bibfield  {title} {\bibinfo {title} {{Curvature Perturbations
  from First-Order Phase Transitions: Implications to Black Holes and
  Gravitational Waves}},\ }\href {https://doi.org/10.1103/tfcx-kzqx} {\bibfield
   {journal} {\bibinfo  {journal} {Phys. Rev. Lett.}\ }\textbf {\bibinfo
  {volume} {136}},\ \bibinfo {pages} {171404} (\bibinfo {year} {2026})},\
  \Eprint {https://arxiv.org/abs/2503.01962} {arXiv:2503.01962 [hep-ph]}
  \BibitemShut {NoStop}%
\bibitem [{\citenamefont {Harada}\ \emph {et~al.}(2015)\citenamefont {Harada},
  \citenamefont {Yoo}, \citenamefont {Nakama},\ and\ \citenamefont
  {Koga}}]{Harada:2015yda}%
  \BibitemOpen
  \bibfield  {author} {\bibinfo {author} {\bibfnamefont {T.}~\bibnamefont
  {Harada}}, \bibinfo {author} {\bibfnamefont {C.-M.}\ \bibnamefont {Yoo}},
  \bibinfo {author} {\bibfnamefont {T.}~\bibnamefont {Nakama}},\ and\ \bibinfo
  {author} {\bibfnamefont {Y.}~\bibnamefont {Koga}},\ }\bibfield  {title}
  {\bibinfo {title} {{Cosmological long-wavelength solutions and primordial
  black hole formation}},\ }\href {https://doi.org/10.1103/PhysRevD.91.084057}
  {\bibfield  {journal} {\bibinfo  {journal} {Phys. Rev. D}\ }\textbf {\bibinfo
  {volume} {91}},\ \bibinfo {pages} {084057} (\bibinfo {year} {2015})},\
  \Eprint {https://arxiv.org/abs/1503.03934} {arXiv:1503.03934 [gr-qc]}
  \BibitemShut {NoStop}%
\bibitem [{\citenamefont {Ai}\ and\ \citenamefont {Xie}(2026)}]{Ai:2026zrs}%
  \BibitemOpen
  \bibfield  {author} {\bibinfo {author} {\bibfnamefont {W.-Y.}\ \bibnamefont
  {Ai}}\ and\ \bibinfo {author} {\bibfnamefont {K.-P.}\ \bibnamefont {Xie}},\
  }\bibfield  {title} {\bibinfo {title} {{Reviving primordial black hole
  formation in slow first-order phase transitions}},\ }\href@noop {} {\
  (\bibinfo {year} {2026})},\ \Eprint {https://arxiv.org/abs/2605.11332}
  {arXiv:2605.11332 [hep-ph]} \BibitemShut {NoStop}%
\bibitem [{\citenamefont {Green}\ \emph {et~al.}(2004)\citenamefont {Green},
  \citenamefont {Liddle}, \citenamefont {Malik},\ and\ \citenamefont
  {Sasaki}}]{Green:2004wb}%
  \BibitemOpen
  \bibfield  {author} {\bibinfo {author} {\bibfnamefont {A.~M.}\ \bibnamefont
  {Green}}, \bibinfo {author} {\bibfnamefont {A.~R.}\ \bibnamefont {Liddle}},
  \bibinfo {author} {\bibfnamefont {K.~A.}\ \bibnamefont {Malik}},\ and\
  \bibinfo {author} {\bibfnamefont {M.}~\bibnamefont {Sasaki}},\ }\bibfield
  {title} {\bibinfo {title} {{A New calculation of the mass fraction of
  primordial black holes}},\ }\href
  {https://doi.org/10.1103/PhysRevD.70.041502} {\bibfield  {journal} {\bibinfo
  {journal} {Phys. Rev. D}\ }\textbf {\bibinfo {volume} {70}},\ \bibinfo
  {pages} {041502} (\bibinfo {year} {2004})},\ \Eprint
  {https://arxiv.org/abs/astro-ph/0403181} {arXiv:astro-ph/0403181}
  \BibitemShut {NoStop}%
\bibitem [{\citenamefont {Musco}\ \emph {et~al.}(2005)\citenamefont {Musco},
  \citenamefont {Miller},\ and\ \citenamefont {Rezzolla}}]{Musco:2004ak}%
  \BibitemOpen
  \bibfield  {author} {\bibinfo {author} {\bibfnamefont {I.}~\bibnamefont
  {Musco}}, \bibinfo {author} {\bibfnamefont {J.~C.}\ \bibnamefont {Miller}},\
  and\ \bibinfo {author} {\bibfnamefont {L.}~\bibnamefont {Rezzolla}},\
  }\bibfield  {title} {\bibinfo {title} {{Computations of primordial black hole
  formation}},\ }\href {https://doi.org/10.1088/0264-9381/22/7/013} {\bibfield
  {journal} {\bibinfo  {journal} {Class. Quant. Grav.}\ }\textbf {\bibinfo
  {volume} {22}},\ \bibinfo {pages} {1405} (\bibinfo {year} {2005})},\ \Eprint
  {https://arxiv.org/abs/gr-qc/0412063} {arXiv:gr-qc/0412063} \BibitemShut
  {NoStop}%
\bibitem [{\citenamefont {Lewicki}\ \emph {et~al.}(2024)\citenamefont
  {Lewicki}, \citenamefont {Toczek},\ and\ \citenamefont
  {Vaskonen}}]{Lewicki:2024ghw}%
  \BibitemOpen
  \bibfield  {author} {\bibinfo {author} {\bibfnamefont {M.}~\bibnamefont
  {Lewicki}}, \bibinfo {author} {\bibfnamefont {P.}~\bibnamefont {Toczek}},\
  and\ \bibinfo {author} {\bibfnamefont {V.}~\bibnamefont {Vaskonen}},\
  }\bibfield  {title} {\bibinfo {title} {{Black Holes and Gravitational Waves
  from Slow First-Order Phase Transitions}},\ }\href
  {https://doi.org/10.1103/PhysRevLett.133.221003} {\bibfield  {journal}
  {\bibinfo  {journal} {Phys. Rev. Lett.}\ }\textbf {\bibinfo {volume} {133}},\
  \bibinfo {pages} {221003} (\bibinfo {year} {2024})},\ \Eprint
  {https://arxiv.org/abs/2402.04158} {arXiv:2402.04158 [astro-ph.CO]}
  \BibitemShut {NoStop}%
\bibitem [{\citenamefont {Lewicki}\ \emph {et~al.}(2025)\citenamefont
  {Lewicki}, \citenamefont {Toczek},\ and\ \citenamefont
  {Vaskonen}}]{Lewicki:2024sfw}%
  \BibitemOpen
  \bibfield  {author} {\bibinfo {author} {\bibfnamefont {M.}~\bibnamefont
  {Lewicki}}, \bibinfo {author} {\bibfnamefont {P.}~\bibnamefont {Toczek}},\
  and\ \bibinfo {author} {\bibfnamefont {V.}~\bibnamefont {Vaskonen}},\
  }\bibfield  {title} {\bibinfo {title} {{Black holes and gravitational waves
  from phase transitions in realistic models}},\ }\href
  {https://doi.org/10.1016/j.dark.2025.102075} {\bibfield  {journal} {\bibinfo
  {journal} {Phys. Dark Univ.}\ }\textbf {\bibinfo {volume} {50}},\ \bibinfo
  {pages} {102075} (\bibinfo {year} {2025})},\ \Eprint
  {https://arxiv.org/abs/2412.10366} {arXiv:2412.10366 [astro-ph.CO]}
  \BibitemShut {NoStop}%
\bibitem [{\citenamefont {Dutta}\ \emph {et~al.}(2026)\citenamefont {Dutta},
  \citenamefont {Hauptmann}, \citenamefont {Huang},\ and\ \citenamefont
  {Thompson}}]{Dutta:2026pbm}%
  \BibitemOpen
  \bibfield  {author} {\bibinfo {author} {\bibfnamefont {B.}~\bibnamefont
  {Dutta}}, \bibinfo {author} {\bibfnamefont {C.}~\bibnamefont {Hauptmann}},
  \bibinfo {author} {\bibfnamefont {P.}~\bibnamefont {Huang}},\ and\ \bibinfo
  {author} {\bibfnamefont {A.}~\bibnamefont {Thompson}},\ }\bibfield  {title}
  {\bibinfo {title} {{PBH formation and Gravitational Waves as Multi-messenger
  Signals of First-order Phase Transitions}},\ }\href@noop {} {\  (\bibinfo
  {year} {2026})},\ \Eprint {https://arxiv.org/abs/2607.15479}
  {arXiv:2607.15479 [hep-ph]} \BibitemShut {NoStop}%
\bibitem [{\citenamefont {Hindmarsh}\ \emph {et~al.}(2015)\citenamefont
  {Hindmarsh}, \citenamefont {Huber}, \citenamefont {Rummukainen},\ and\
  \citenamefont {Weir}}]{Hindmarsh:2015qta}%
  \BibitemOpen
  \bibfield  {author} {\bibinfo {author} {\bibfnamefont {M.}~\bibnamefont
  {Hindmarsh}}, \bibinfo {author} {\bibfnamefont {S.~J.}\ \bibnamefont
  {Huber}}, \bibinfo {author} {\bibfnamefont {K.}~\bibnamefont {Rummukainen}},\
  and\ \bibinfo {author} {\bibfnamefont {D.~J.}\ \bibnamefont {Weir}},\
  }\bibfield  {title} {\bibinfo {title} {{Numerical simulations of acoustically
  generated gravitational waves at a first order phase transition}},\ }\href
  {https://doi.org/10.1103/PhysRevD.92.123009} {\bibfield  {journal} {\bibinfo
  {journal} {Phys. Rev. D}\ }\textbf {\bibinfo {volume} {92}},\ \bibinfo
  {pages} {123009} (\bibinfo {year} {2015})},\ \Eprint
  {https://arxiv.org/abs/1504.03291} {arXiv:1504.03291 [astro-ph.CO]}
  \BibitemShut {NoStop}%
\bibitem [{\citenamefont {Hindmarsh}\ \emph {et~al.}(2014)\citenamefont
  {Hindmarsh}, \citenamefont {Huber}, \citenamefont {Rummukainen},\ and\
  \citenamefont {Weir}}]{Hindmarsh:2013xza}%
  \BibitemOpen
  \bibfield  {author} {\bibinfo {author} {\bibfnamefont {M.}~\bibnamefont
  {Hindmarsh}}, \bibinfo {author} {\bibfnamefont {S.~J.}\ \bibnamefont
  {Huber}}, \bibinfo {author} {\bibfnamefont {K.}~\bibnamefont {Rummukainen}},\
  and\ \bibinfo {author} {\bibfnamefont {D.~J.}\ \bibnamefont {Weir}},\
  }\bibfield  {title} {\bibinfo {title} {{Gravitational waves from the sound of
  a first order phase transition}},\ }\href
  {https://doi.org/10.1103/PhysRevLett.112.041301} {\bibfield  {journal}
  {\bibinfo  {journal} {Phys. Rev. Lett.}\ }\textbf {\bibinfo {volume} {112}},\
  \bibinfo {pages} {041301} (\bibinfo {year} {2014})},\ \Eprint
  {https://arxiv.org/abs/1304.2433} {arXiv:1304.2433 [hep-ph]} \BibitemShut
  {NoStop}%
\bibitem [{\citenamefont {Ellis}\ \emph
  {et~al.}(2019{\natexlab{a}})\citenamefont {Ellis}, \citenamefont {Lewicki},\
  and\ \citenamefont {No}}]{Ellis:2018mja}%
  \BibitemOpen
  \bibfield  {author} {\bibinfo {author} {\bibfnamefont {J.}~\bibnamefont
  {Ellis}}, \bibinfo {author} {\bibfnamefont {M.}~\bibnamefont {Lewicki}},\
  and\ \bibinfo {author} {\bibfnamefont {J.~M.}\ \bibnamefont {No}},\
  }\bibfield  {title} {\bibinfo {title} {{On the Maximal Strength of a
  First-Order Electroweak Phase Transition and its Gravitational Wave
  Signal}},\ }\href {https://doi.org/10.1088/1475-7516/2019/04/003} {\bibfield
  {journal} {\bibinfo  {journal} {JCAP}\ }\textbf {\bibinfo {volume} {04}},\
  \bibinfo {pages} {003}},\ \Eprint {https://arxiv.org/abs/1809.08242}
  {arXiv:1809.08242 [hep-ph]} \BibitemShut {NoStop}%
\bibitem [{\citenamefont {Costa}\ \emph {et~al.}(2025)\citenamefont {Costa},
  \citenamefont {Hoefken~Zink}, \citenamefont {Lucente}, \citenamefont
  {Pascoli},\ and\ \citenamefont {Rosauro-Alcaraz}}]{Costa:2025csj}%
  \BibitemOpen
  \bibfield  {author} {\bibinfo {author} {\bibfnamefont {F.}~\bibnamefont
  {Costa}}, \bibinfo {author} {\bibfnamefont {J.}~\bibnamefont {Hoefken~Zink}},
  \bibinfo {author} {\bibfnamefont {M.}~\bibnamefont {Lucente}}, \bibinfo
  {author} {\bibfnamefont {S.}~\bibnamefont {Pascoli}},\ and\ \bibinfo {author}
  {\bibfnamefont {S.}~\bibnamefont {Rosauro-Alcaraz}},\ }\bibfield  {title}
  {\bibinfo {title} {{Supercooled dark scalar phase transitions explanation of
  NANOGrav data}},\ }\href {https://doi.org/10.1016/j.physletb.2025.139634}
  {\bibfield  {journal} {\bibinfo  {journal} {Phys. Lett. B}\ }\textbf
  {\bibinfo {volume} {868}},\ \bibinfo {pages} {139634} (\bibinfo {year}
  {2025})},\ \Eprint {https://arxiv.org/abs/2501.15649} {arXiv:2501.15649
  [hep-ph]} \BibitemShut {NoStop}%
\bibitem [{\citenamefont {Gouttenoire}\ \emph {et~al.}(2022)\citenamefont
  {Gouttenoire}, \citenamefont {Jinno},\ and\ \citenamefont
  {Sala}}]{Gouttenoire:2021kjv}%
  \BibitemOpen
  \bibfield  {author} {\bibinfo {author} {\bibfnamefont {Y.}~\bibnamefont
  {Gouttenoire}}, \bibinfo {author} {\bibfnamefont {R.}~\bibnamefont {Jinno}},\
  and\ \bibinfo {author} {\bibfnamefont {F.}~\bibnamefont {Sala}},\ }\bibfield
  {title} {\bibinfo {title} {{Friction pressure on relativistic bubble
  walls}},\ }\href {https://doi.org/10.1007/JHEP05(2022)004} {\bibfield
  {journal} {\bibinfo  {journal} {JHEP}\ }\textbf {\bibinfo {volume} {05}},\
  \bibinfo {pages} {004}},\ \Eprint {https://arxiv.org/abs/2112.07686}
  {arXiv:2112.07686 [hep-ph]} \BibitemShut {NoStop}%
\bibitem [{\citenamefont {Bodeker}\ and\ \citenamefont
  {Moore}(2009)}]{Bodeker:2009qy}%
  \BibitemOpen
  \bibfield  {author} {\bibinfo {author} {\bibfnamefont {D.}~\bibnamefont
  {Bodeker}}\ and\ \bibinfo {author} {\bibfnamefont {G.~D.}\ \bibnamefont
  {Moore}},\ }\bibfield  {title} {\bibinfo {title} {{Can electroweak bubble
  walls run away?}},\ }\href {https://doi.org/10.1088/1475-7516/2009/05/009}
  {\bibfield  {journal} {\bibinfo  {journal} {JCAP}\ }\textbf {\bibinfo
  {volume} {05}},\ \bibinfo {pages} {009}},\ \Eprint
  {https://arxiv.org/abs/0903.4099} {arXiv:0903.4099 [hep-ph]} \BibitemShut
  {NoStop}%
\bibitem [{\citenamefont {Bodeker}\ and\ \citenamefont
  {Moore}(2017)}]{Bodeker:2017cim}%
  \BibitemOpen
  \bibfield  {author} {\bibinfo {author} {\bibfnamefont {D.}~\bibnamefont
  {Bodeker}}\ and\ \bibinfo {author} {\bibfnamefont {G.~D.}\ \bibnamefont
  {Moore}},\ }\bibfield  {title} {\bibinfo {title} {{Electroweak Bubble Wall
  Speed Limit}},\ }\href {https://doi.org/10.1088/1475-7516/2017/05/025}
  {\bibfield  {journal} {\bibinfo  {journal} {JCAP}\ }\textbf {\bibinfo
  {volume} {05}},\ \bibinfo {pages} {025}},\ \Eprint
  {https://arxiv.org/abs/1703.08215} {arXiv:1703.08215 [hep-ph]} \BibitemShut
  {NoStop}%
\bibitem [{\citenamefont {Barroso~Mancha}\ \emph {et~al.}(2021)\citenamefont
  {Barroso~Mancha}, \citenamefont {Prokopec},\ and\ \citenamefont
  {Swiezewska}}]{BarrosoMancha:2020fay}%
  \BibitemOpen
  \bibfield  {author} {\bibinfo {author} {\bibfnamefont {M.}~\bibnamefont
  {Barroso~Mancha}}, \bibinfo {author} {\bibfnamefont {T.}~\bibnamefont
  {Prokopec}},\ and\ \bibinfo {author} {\bibfnamefont {B.}~\bibnamefont
  {Swiezewska}},\ }\bibfield  {title} {\bibinfo {title} {{Field-theoretic
  derivation of bubble-wall force}},\ }\href
  {https://doi.org/10.1007/JHEP01(2021)070} {\bibfield  {journal} {\bibinfo
  {journal} {JHEP}\ }\textbf {\bibinfo {volume} {01}},\ \bibinfo {pages}
  {070}},\ \Eprint {https://arxiv.org/abs/2005.10875} {arXiv:2005.10875
  [hep-th]} \BibitemShut {NoStop}%
\bibitem [{\citenamefont {Long}\ and\ \citenamefont
  {Turner}(2024)}]{Long:2024sqg}%
  \BibitemOpen
  \bibfield  {author} {\bibinfo {author} {\bibfnamefont {A.~J.}\ \bibnamefont
  {Long}}\ and\ \bibinfo {author} {\bibfnamefont {J.}~\bibnamefont {Turner}},\
  }\bibfield  {title} {\bibinfo {title} {{Thermal pressure on ultrarelativistic
  bubbles from a semiclassical formalism}},\ }\href
  {https://doi.org/10.1088/1475-7516/2024/11/024} {\bibfield  {journal}
  {\bibinfo  {journal} {JCAP}\ }\textbf {\bibinfo {volume} {11}},\ \bibinfo
  {pages} {024}},\ \Eprint {https://arxiv.org/abs/2407.18196} {arXiv:2407.18196
  [hep-ph]} \BibitemShut {NoStop}%
\bibitem [{\citenamefont {H{\"o}che}\ \emph {et~al.}(2021)\citenamefont
  {H{\"o}che}, \citenamefont {Kozaczuk}, \citenamefont {Long}, \citenamefont
  {Turner},\ and\ \citenamefont {Wang}}]{Hoche:2020ysm}%
  \BibitemOpen
  \bibfield  {author} {\bibinfo {author} {\bibfnamefont {S.}~\bibnamefont
  {H{\"o}che}}, \bibinfo {author} {\bibfnamefont {J.}~\bibnamefont {Kozaczuk}},
  \bibinfo {author} {\bibfnamefont {A.~J.}\ \bibnamefont {Long}}, \bibinfo
  {author} {\bibfnamefont {J.}~\bibnamefont {Turner}},\ and\ \bibinfo {author}
  {\bibfnamefont {Y.}~\bibnamefont {Wang}},\ }\bibfield  {title} {\bibinfo
  {title} {{Towards an all-orders calculation of the electroweak bubble wall
  velocity}},\ }\href {https://doi.org/10.1088/1475-7516/2021/03/009}
  {\bibfield  {journal} {\bibinfo  {journal} {JCAP}\ }\textbf {\bibinfo
  {volume} {03}},\ \bibinfo {pages} {009}},\ \Eprint
  {https://arxiv.org/abs/2007.10343} {arXiv:2007.10343 [hep-ph]} \BibitemShut
  {NoStop}%
\bibitem [{\citenamefont {Gon{\c{c}}alves}\ \emph {et~al.}(2025)\citenamefont
  {Gon{\c{c}}alves}, \citenamefont {Kaladharan},\ and\ \citenamefont
  {Wu}}]{Goncalves:2024vkj}%
  \BibitemOpen
  \bibfield  {author} {\bibinfo {author} {\bibfnamefont {D.}~\bibnamefont
  {Gon{\c{c}}alves}}, \bibinfo {author} {\bibfnamefont {A.}~\bibnamefont
  {Kaladharan}},\ and\ \bibinfo {author} {\bibfnamefont {Y.}~\bibnamefont
  {Wu}},\ }\bibfield  {title} {\bibinfo {title} {{Primordial black holes from
  first-order phase transition in the singlet-extended SM}},\ }\href
  {https://doi.org/10.1103/PhysRevD.111.035009} {\bibfield  {journal} {\bibinfo
   {journal} {Phys. Rev. D}\ }\textbf {\bibinfo {volume} {111}},\ \bibinfo
  {pages} {035009} (\bibinfo {year} {2025})},\ \Eprint
  {https://arxiv.org/abs/2406.07622} {arXiv:2406.07622 [hep-ph]} \BibitemShut
  {NoStop}%
\bibitem [{\citenamefont {Cutting}\ \emph {et~al.}(2018)\citenamefont
  {Cutting}, \citenamefont {Hindmarsh},\ and\ \citenamefont
  {Weir}}]{Cutting:2018tjt}%
  \BibitemOpen
  \bibfield  {author} {\bibinfo {author} {\bibfnamefont {D.}~\bibnamefont
  {Cutting}}, \bibinfo {author} {\bibfnamefont {M.}~\bibnamefont {Hindmarsh}},\
  and\ \bibinfo {author} {\bibfnamefont {D.~J.}\ \bibnamefont {Weir}},\
  }\bibfield  {title} {\bibinfo {title} {{Gravitational waves from vacuum
  first-order phase transitions: from the envelope to the lattice}},\ }\href
  {https://doi.org/10.1103/PhysRevD.97.123513} {\bibfield  {journal} {\bibinfo
  {journal} {Phys. Rev. D}\ }\textbf {\bibinfo {volume} {97}},\ \bibinfo
  {pages} {123513} (\bibinfo {year} {2018})},\ \Eprint
  {https://arxiv.org/abs/1802.05712} {arXiv:1802.05712 [astro-ph.CO]}
  \BibitemShut {NoStop}%
\bibitem [{\citenamefont {Antoniadis}\ \emph {et~al.}(2022)\citenamefont
  {Antoniadis} \emph {et~al.}}]{Antoniadis:2022pcn}%
  \BibitemOpen
  \bibfield  {author} {\bibinfo {author} {\bibfnamefont {J.}~\bibnamefont
  {Antoniadis}} \emph {et~al.},\ }\bibfield  {title} {\bibinfo {title} {{The
  International Pulsar Timing Array second data release: Search for an
  isotropic gravitational wave background}},\ }\href
  {https://doi.org/10.1093/mnras/stab3418} {\bibfield  {journal} {\bibinfo
  {journal} {Mon. Not. Roy. Astron. Soc.}\ }\textbf {\bibinfo {volume} {510}},\
  \bibinfo {pages} {4873} (\bibinfo {year} {2022})},\ \Eprint
  {https://arxiv.org/abs/2201.03980} {arXiv:2201.03980 [astro-ph.HE]}
  \BibitemShut {NoStop}%
\bibitem [{\citenamefont {Abbott}\ \emph
  {et~al.}(2017{\natexlab{b}})\citenamefont {Abbott} \emph
  {et~al.}}]{LIGOScientific:2016wof}%
  \BibitemOpen
  \bibfield  {author} {\bibinfo {author} {\bibfnamefont {B.~P.}\ \bibnamefont
  {Abbott}} \emph {et~al.} (\bibinfo {collaboration} {LIGO Scientific}),\
  }\bibfield  {title} {\bibinfo {title} {{Exploring the Sensitivity of Next
  Generation Gravitational Wave Detectors}},\ }\href
  {https://doi.org/10.1088/1361-6382/aa51f4} {\bibfield  {journal} {\bibinfo
  {journal} {Class. Quant. Grav.}\ }\textbf {\bibinfo {volume} {34}},\ \bibinfo
  {pages} {044001} (\bibinfo {year} {2017}{\natexlab{b}})},\ \Eprint
  {https://arxiv.org/abs/1607.08697} {arXiv:1607.08697 [astro-ph.IM]}
  \BibitemShut {NoStop}%
\bibitem [{\citenamefont {Janssen}\ \emph {et~al.}(2015)\citenamefont {Janssen}
  \emph {et~al.}}]{Janssen:2014dka}%
  \BibitemOpen
  \bibfield  {author} {\bibinfo {author} {\bibfnamefont {G.}~\bibnamefont
  {Janssen}} \emph {et~al.},\ }\bibfield  {title} {\bibinfo {title}
  {{Gravitational wave astronomy with the SKA}},\ }\href
  {https://doi.org/10.22323/1.215.0037} {\bibfield  {journal} {\bibinfo
  {journal} {PoS}\ }\textbf {\bibinfo {volume} {AASKA14}},\ \bibinfo {pages}
  {037} (\bibinfo {year} {2015})},\ \Eprint {https://arxiv.org/abs/1501.00127}
  {arXiv:1501.00127 [astro-ph.IM]} \BibitemShut {NoStop}%
\bibitem [{\citenamefont {Amaro-Seoane}\ \emph {et~al.}(2017)\citenamefont
  {Amaro-Seoane} \emph {et~al.}}]{LISA:2017pwj}%
  \BibitemOpen
  \bibfield  {author} {\bibinfo {author} {\bibfnamefont {P.}~\bibnamefont
  {Amaro-Seoane}} \emph {et~al.} (\bibinfo {collaboration} {LISA}),\ }\bibfield
   {title} {\bibinfo {title} {{Laser Interferometer Space Antenna}},\
  }\href@noop {} {\  (\bibinfo {year} {2017})},\ \Eprint
  {https://arxiv.org/abs/1702.00786} {arXiv:1702.00786 [astro-ph.IM]}
  \BibitemShut {NoStop}%
\bibitem [{\citenamefont {Sesana}\ \emph {et~al.}(2021)\citenamefont {Sesana}
  \emph {et~al.}}]{Sesana:2019vho}%
  \BibitemOpen
  \bibfield  {author} {\bibinfo {author} {\bibfnamefont {A.}~\bibnamefont
  {Sesana}} \emph {et~al.},\ }\bibfield  {title} {\bibinfo {title} {{Unveiling
  the gravitational universe at $\mu$-Hz frequencies}},\ }\href
  {https://doi.org/10.1007/s10686-021-09709-9} {\bibfield  {journal} {\bibinfo
  {journal} {Exper. Astron.}\ }\textbf {\bibinfo {volume} {51}},\ \bibinfo
  {pages} {1333} (\bibinfo {year} {2021})},\ \Eprint
  {https://arxiv.org/abs/1908.11391} {arXiv:1908.11391 [astro-ph.IM]}
  \BibitemShut {NoStop}%
\bibitem [{\citenamefont {El-Neaj}\ \emph {et~al.}(2020)\citenamefont {El-Neaj}
  \emph {et~al.}}]{AEDGE:2019nxb}%
  \BibitemOpen
  \bibfield  {author} {\bibinfo {author} {\bibfnamefont {Y.~A.}\ \bibnamefont
  {El-Neaj}} \emph {et~al.} (\bibinfo {collaboration} {AEDGE}),\ }\bibfield
  {title} {\bibinfo {title} {{AEDGE: Atomic Experiment for Dark Matter and
  Gravity Exploration in Space}},\ }\href
  {https://doi.org/10.1140/epjqt/s40507-020-0080-0} {\bibfield  {journal}
  {\bibinfo  {journal} {EPJ Quant. Technol.}\ }\textbf {\bibinfo {volume}
  {7}},\ \bibinfo {pages} {6} (\bibinfo {year} {2020})},\ \Eprint
  {https://arxiv.org/abs/1908.00802} {arXiv:1908.00802 [gr-qc]} \BibitemShut
  {NoStop}%
\bibitem [{\citenamefont {Badurina}\ \emph {et~al.}(2020)\citenamefont
  {Badurina} \emph {et~al.}}]{Badurina:2019hst}%
  \BibitemOpen
  \bibfield  {author} {\bibinfo {author} {\bibfnamefont {L.}~\bibnamefont
  {Badurina}} \emph {et~al.},\ }\bibfield  {title} {\bibinfo {title} {{AION: An
  Atom Interferometer Observatory and Network}},\ }\href
  {https://doi.org/10.1088/1475-7516/2020/05/011} {\bibfield  {journal}
  {\bibinfo  {journal} {JCAP}\ }\textbf {\bibinfo {volume} {05}},\ \bibinfo
  {pages} {011}},\ \Eprint {https://arxiv.org/abs/1911.11755} {arXiv:1911.11755
  [astro-ph.CO]} \BibitemShut {NoStop}%
\bibitem [{\citenamefont {Kawamura}\ \emph {et~al.}(2021)\citenamefont
  {Kawamura} \emph {et~al.}}]{Kawamura:2020pcg}%
  \BibitemOpen
  \bibfield  {author} {\bibinfo {author} {\bibfnamefont {S.}~\bibnamefont
  {Kawamura}} \emph {et~al.},\ }\bibfield  {title} {\bibinfo {title} {{Current
  status of space gravitational wave antenna DECIGO and B-DECIGO}},\ }\href
  {https://doi.org/10.1093/ptep/ptab019} {\bibfield  {journal} {\bibinfo
  {journal} {PTEP}\ }\textbf {\bibinfo {volume} {2021}},\ \bibinfo {pages}
  {05A105} (\bibinfo {year} {2021})},\ \Eprint
  {https://arxiv.org/abs/2006.13545} {arXiv:2006.13545 [gr-qc]} \BibitemShut
  {NoStop}%
\bibitem [{\citenamefont {Kudoh}\ \emph {et~al.}(2006)\citenamefont {Kudoh}
  \emph {et~al.}}]{Kudoh:2005as}%
  \BibitemOpen
  \bibfield  {author} {\bibinfo {author} {\bibfnamefont {H.}~\bibnamefont
  {Kudoh}} \emph {et~al.},\ }\bibfield  {title} {\bibinfo {title} {{Detecting a
  gravitational-wave background with next-generation space interferometers}},\
  }\href {https://doi.org/10.1103/PhysRevD.73.064006} {\bibfield  {journal}
  {\bibinfo  {journal} {Phys. Rev. D}\ }\textbf {\bibinfo {volume} {73}},\
  \bibinfo {pages} {064006} (\bibinfo {year} {2006})},\ \Eprint
  {https://arxiv.org/abs/gr-qc/0511145} {arXiv:gr-qc/0511145} \BibitemShut
  {NoStop}%
\bibitem [{\citenamefont {Harry}\ \emph {et~al.}(2006)\citenamefont {Harry},
  \citenamefont {Fritschel}, \citenamefont {Shaddock}, \citenamefont
  {Folkner},\ and\ \citenamefont {Phinney}}]{Harry:2006fi}%
  \BibitemOpen
  \bibfield  {author} {\bibinfo {author} {\bibfnamefont {G.~M.}\ \bibnamefont
  {Harry}}, \bibinfo {author} {\bibfnamefont {P.}~\bibnamefont {Fritschel}},
  \bibinfo {author} {\bibfnamefont {D.~A.}\ \bibnamefont {Shaddock}}, \bibinfo
  {author} {\bibfnamefont {W.}~\bibnamefont {Folkner}},\ and\ \bibinfo {author}
  {\bibfnamefont {E.~S.}\ \bibnamefont {Phinney}},\ }\bibfield  {title}
  {\bibinfo {title} {{Laser interferometry for the big bang observer}},\ }\href
  {https://doi.org/10.1088/0264-9381/23/15/008} {\bibfield  {journal} {\bibinfo
   {journal} {Class. Quant. Grav.}\ }\textbf {\bibinfo {volume} {23}},\
  \bibinfo {pages} {4887} (\bibinfo {year} {2006})},\ \bibinfo {note}
  {[Erratum: Class.Quant.Grav. 23, 7361 (2006)]}\BibitemShut {NoStop}%
\bibitem [{\citenamefont {Hild}\ \emph {et~al.}(2008)\citenamefont {Hild},
  \citenamefont {Chelkowski},\ and\ \citenamefont {Freise}}]{Hild:2008ng}%
  \BibitemOpen
  \bibfield  {author} {\bibinfo {author} {\bibfnamefont {S.}~\bibnamefont
  {Hild}}, \bibinfo {author} {\bibfnamefont {S.}~\bibnamefont {Chelkowski}},\
  and\ \bibinfo {author} {\bibfnamefont {A.}~\bibnamefont {Freise}},\
  }\bibfield  {title} {\bibinfo {title} {{Pushing towards the ET sensitivity
  using 'conventional' technology}},\ }\href@noop {} {\  (\bibinfo {year}
  {2008})},\ \Eprint {https://arxiv.org/abs/0810.0604} {arXiv:0810.0604
  [gr-qc]} \BibitemShut {NoStop}%
\bibitem [{\citenamefont {Espinosa}\ \emph {et~al.}(2010)\citenamefont
  {Espinosa}, \citenamefont {Konstandin}, \citenamefont {No},\ and\
  \citenamefont {Servant}}]{Espinosa:2010hh}%
  \BibitemOpen
  \bibfield  {author} {\bibinfo {author} {\bibfnamefont {J.~R.}\ \bibnamefont
  {Espinosa}}, \bibinfo {author} {\bibfnamefont {T.}~\bibnamefont
  {Konstandin}}, \bibinfo {author} {\bibfnamefont {J.~M.}\ \bibnamefont {No}},\
  and\ \bibinfo {author} {\bibfnamefont {G.}~\bibnamefont {Servant}},\
  }\bibfield  {title} {\bibinfo {title} {{Energy Budget of Cosmological
  First-order Phase Transitions}},\ }\href
  {https://doi.org/10.1088/1475-7516/2010/06/028} {\bibfield  {journal}
  {\bibinfo  {journal} {JCAP}\ }\textbf {\bibinfo {volume} {06}},\ \bibinfo
  {pages} {028}},\ \Eprint {https://arxiv.org/abs/1004.4187} {arXiv:1004.4187
  [hep-ph]} \BibitemShut {NoStop}%
\bibitem [{\citenamefont {Caprini}\ \emph {et~al.}(2016)\citenamefont {Caprini}
  \emph {et~al.}}]{Caprini:2015zlo}%
  \BibitemOpen
  \bibfield  {author} {\bibinfo {author} {\bibfnamefont {C.}~\bibnamefont
  {Caprini}} \emph {et~al.},\ }\bibfield  {title} {\bibinfo {title} {{Science
  with the space-based interferometer eLISA. II: Gravitational waves from
  cosmological phase transitions}},\ }\href
  {https://doi.org/10.1088/1475-7516/2016/04/001} {\bibfield  {journal}
  {\bibinfo  {journal} {JCAP}\ }\textbf {\bibinfo {volume} {04}},\ \bibinfo
  {pages} {001}},\ \Eprint {https://arxiv.org/abs/1512.06239} {arXiv:1512.06239
  [astro-ph.CO]} \BibitemShut {NoStop}%
\bibitem [{\citenamefont {Jana}\ \emph {et~al.}(2026)\citenamefont {Jana},
  \citenamefont {Manna},\ and\ \citenamefont {K}}]{Jana:2025vyb}%
  \BibitemOpen
  \bibfield  {author} {\bibinfo {author} {\bibfnamefont {S.}~\bibnamefont
  {Jana}}, \bibinfo {author} {\bibfnamefont {S.}~\bibnamefont {Manna}},\ and\
  \bibinfo {author} {\bibfnamefont {V.~P.}\ \bibnamefont {K}},\ }\bibfield
  {title} {\bibinfo {title} {{Gravitational Wave Signature and the Nature of
  Neutrino Masses: Majorana, Dirac, or Pseudo-Dirac?}},\ }\href
  {https://doi.org/10.1016/j.physletb.2026.140476} {\bibfield  {journal}
  {\bibinfo  {journal} {Phys. Lett. B}\ }\textbf {\bibinfo {volume} {877}},\
  \bibinfo {pages} {140476} (\bibinfo {year} {2026})},\ \Eprint
  {https://arxiv.org/abs/2509.10456} {arXiv:2509.10456 [hep-ph]} \BibitemShut
  {NoStop}%
\bibitem [{\citenamefont {Kamionkowski}\ \emph {et~al.}(1994)\citenamefont
  {Kamionkowski}, \citenamefont {Kosowsky},\ and\ \citenamefont
  {Turner}}]{Kamionkowski:1993fg}%
  \BibitemOpen
  \bibfield  {author} {\bibinfo {author} {\bibfnamefont {M.}~\bibnamefont
  {Kamionkowski}}, \bibinfo {author} {\bibfnamefont {A.}~\bibnamefont
  {Kosowsky}},\ and\ \bibinfo {author} {\bibfnamefont {M.~S.}\ \bibnamefont
  {Turner}},\ }\bibfield  {title} {\bibinfo {title} {{Gravitational radiation
  from first order phase transitions}},\ }\href
  {https://doi.org/10.1103/PhysRevD.49.2837} {\bibfield  {journal} {\bibinfo
  {journal} {Phys. Rev. D}\ }\textbf {\bibinfo {volume} {49}},\ \bibinfo
  {pages} {2837} (\bibinfo {year} {1994})},\ \Eprint
  {https://arxiv.org/abs/astro-ph/9310044} {arXiv:astro-ph/9310044}
  \BibitemShut {NoStop}%
\bibitem [{\citenamefont {Caprini}\ \emph {et~al.}(2009)\citenamefont
  {Caprini}, \citenamefont {Durrer},\ and\ \citenamefont
  {Servant}}]{Caprini:2009yp}%
  \BibitemOpen
  \bibfield  {author} {\bibinfo {author} {\bibfnamefont {C.}~\bibnamefont
  {Caprini}}, \bibinfo {author} {\bibfnamefont {R.}~\bibnamefont {Durrer}},\
  and\ \bibinfo {author} {\bibfnamefont {G.}~\bibnamefont {Servant}},\
  }\bibfield  {title} {\bibinfo {title} {{The stochastic gravitational wave
  background from turbulence and magnetic fields generated by a first-order
  phase transition}},\ }\href {https://doi.org/10.1088/1475-7516/2009/12/024}
  {\bibfield  {journal} {\bibinfo  {journal} {JCAP}\ }\textbf {\bibinfo
  {volume} {12}},\ \bibinfo {pages} {024}},\ \Eprint
  {https://arxiv.org/abs/0909.0622} {arXiv:0909.0622 [astro-ph.CO]}
  \BibitemShut {NoStop}%
\bibitem [{\citenamefont {Jinno}\ \emph {et~al.}(2016)\citenamefont {Jinno},
  \citenamefont {Nakayama},\ and\ \citenamefont {Takimoto}}]{Jinno:2015doa}%
  \BibitemOpen
  \bibfield  {author} {\bibinfo {author} {\bibfnamefont {R.}~\bibnamefont
  {Jinno}}, \bibinfo {author} {\bibfnamefont {K.}~\bibnamefont {Nakayama}},\
  and\ \bibinfo {author} {\bibfnamefont {M.}~\bibnamefont {Takimoto}},\
  }\bibfield  {title} {\bibinfo {title} {{Gravitational waves from the first
  order phase transition of the Higgs field at high energy scales}},\ }\href
  {https://doi.org/10.1103/PhysRevD.93.045024} {\bibfield  {journal} {\bibinfo
  {journal} {Phys. Rev. D}\ }\textbf {\bibinfo {volume} {93}},\ \bibinfo
  {pages} {045024} (\bibinfo {year} {2016})},\ \Eprint
  {https://arxiv.org/abs/1510.02697} {arXiv:1510.02697 [hep-ph]} \BibitemShut
  {NoStop}%
\bibitem [{\citenamefont {Ellis}\ \emph
  {et~al.}(2019{\natexlab{b}})\citenamefont {Ellis}, \citenamefont {Lewicki},
  \citenamefont {No},\ and\ \citenamefont {Vaskonen}}]{Ellis:2019oqb}%
  \BibitemOpen
  \bibfield  {author} {\bibinfo {author} {\bibfnamefont {J.}~\bibnamefont
  {Ellis}}, \bibinfo {author} {\bibfnamefont {M.}~\bibnamefont {Lewicki}},
  \bibinfo {author} {\bibfnamefont {J.~M.}\ \bibnamefont {No}},\ and\ \bibinfo
  {author} {\bibfnamefont {V.}~\bibnamefont {Vaskonen}},\ }\bibfield  {title}
  {\bibinfo {title} {{Gravitational wave energy budget in strongly supercooled
  phase transitions}},\ }\href {https://doi.org/10.1088/1475-7516/2019/06/024}
  {\bibfield  {journal} {\bibinfo  {journal} {JCAP}\ }\textbf {\bibinfo
  {volume} {06}},\ \bibinfo {pages} {024}},\ \Eprint
  {https://arxiv.org/abs/1903.09642} {arXiv:1903.09642 [hep-ph]} \BibitemShut
  {NoStop}%
\bibitem [{\citenamefont {Laurent}\ and\ \citenamefont
  {Vanvlasselaer}(2026)}]{Laurent:2026jvx}%
  \BibitemOpen
  \bibfield  {author} {\bibinfo {author} {\bibfnamefont {B.}~\bibnamefont
  {Laurent}}\ and\ \bibinfo {author} {\bibfnamefont {M.}~\bibnamefont
  {Vanvlasselaer}},\ }\bibfield  {title} {\bibinfo {title} {{Dynamical
  evolution of the pressure on the bubble wall}},\ }\href@noop {} {\  (\bibinfo
  {year} {2026})},\ \Eprint {https://arxiv.org/abs/2606.30740}
  {arXiv:2606.30740 [hep-ph]} \BibitemShut {NoStop}%
\end{thebibliography}%

\end{document}